\documentclass[journal,twocolumn,10pt,final,letterpaper]{IEEEtran}

\usepackage{setspace}
\usepackage{cite}
\usepackage{hyperref}
\ifCLASSINFOpdf
  \usepackage[pdftex]{graphicx}
  \DeclareGraphicsExtensions{.pdf,.jpeg,.png}
\else
   \usepackage[dvips]{graphicx}
   \DeclareGraphicsExtensions{.eps}
\fi

\usepackage[cmex10]{amsmath}
\usepackage{amssymb,amsthm}
\usepackage{mathtools}
\usepackage{algorithmicx}
\usepackage{algpseudocode}
\newtheorem{remark}{Remark}

\newtheorem{theorem}{Theorem}
\newtheorem{corollary}{Corollary}
\newtheorem{definition}{Definition}

\newtheorem{proposition}{Proposition}

\newcommand{\hypergeombf}[4]{{}_{2}\hspace{-.01cm}\mathbf{F}_{\hspace{-.04cm}1}\hspace{-.1cm}\left(#1,#2,#3;#4\right)}
\newcommand{\cp}{\mathtt{P}_{\text{\textnormal{MRC}}}}
\newcommand{\sinr}{\mathtt{SINR}}
\newcommand{\sinrmrc}{\sinr_{\text{\textnormal{MRC}}}}
\renewcommand{\th}{^\text{th}}
\newcommand{\hyperg}{{}_{2}F_{1}}
\newcommand{\cpfc}{\mathtt{P}^{\text{\textnormal{FC}}}_{\text{\textnormal{MRC}}}}
\newcommand{\cpnc}{\mathtt{P}^{\text{\textnormal{NC}}}_{\text{\textnormal{MRC}}}}
\newcommand{\cpsc}{\mathtt{P}_{\text{\textnormal{SC}}}}
\newcommand{\cpmmse}{\mathtt{P}_{\text{\textnormal{MMSE}}}}
\newcommand{\snr}{\mathtt{SNR}}
\newcommand{\mathdef}{~\raisebox{-0.03cm}{$\triangleq$}~}
\newcommand{\sinrfc}{\mathtt{SINR}^\text{\textnormal{FC}}_{\text{\textnormal{MRC}}}}
\newcommand{\sinrnc}{\mathtt{SINR}^\text{\textnormal{NC}}_{\text{\textnormal{MRC}}}}
\newcommand{\md}{m_{\text{\textnormal{D}}}}
\newcommand{\mi}{m_{\text{\textnormal{I}}}}

\newsavebox{\ieeealgbox}
\newenvironment{boxedalgorithmic}
  {\begin{lrbox}{\ieeealgbox}~\hspace{-.25cm}
   \begin{minipage}{.475\textwidth}
   \begin{algorithmic}[1]}
  {\end{algorithmic}
   \end{minipage}
   \end{lrbox}\noindent\fbox{\usebox{\ieeealgbox}}}

\algblockdefx{ForP}{EndForP}[1]%
  {\textbf{for} #1 \textbf{do in parallel}}%
  {\textbf{end for}}
  
\algblockdefx{Get}{EndForP}[1]%
  {\textbf{for} #1 \textbf{do in parallel}}%
  {\textbf{end for}}

\usepackage{psfrag}
\usepackage{array}
\usepackage{mdwmath}
\usepackage{mdwtab}
\usepackage{enumerate}
\usepackage{cases}

\let\underbrace\LaTeXunderbrace

\usepackage[caption=false,font=footnotesize,labelfont=sf,textfont=sf]{subfig}

\usepackage{fixltx2e}
\usepackage{url}

\newcounter{tmp_equation}
\allowdisplaybreaks

\begin{document}

\title{Dual-Branch MRC Receivers under Spatial Interference Correlation and Nakagami Fading}

\author{Ralph Tanbourgi\IEEEauthorrefmark{1},~\IEEEmembership{Student Member,~IEEE,} Harpreet S. Dhillon\IEEEauthorrefmark{2},~\IEEEmembership{Member,~IEEE,}\\Jeffrey G. Andrews\IEEEauthorrefmark{3},~\IEEEmembership{Fellow,~IEEE,} and
Friedrich K. Jondral\IEEEauthorrefmark{1},~\IEEEmembership{Senior Member,~IEEE}

\thanks{\IEEEauthorrefmark{1}R.~Tanbourgi and F.~K.~Jondral are with the Communications Engineering Lab (CEL), Karlsruhe Institute of Technology (KIT), Germany. Email: \texttt{\{ralph.tanbourgi, friedrich.jondral\}@kit.edu}. This work was partially supported by the German Research Foundation (DFG) within the Priority Program 1397 "COIN" under grant No. JO258/21-1 and JO258/21-2.}
\thanks{\IEEEauthorrefmark{2}H. S. Dhillon is with the Communication Sciences Institute (CSI), Department of Electrical Engineering, University of Southern California, Los Angeles, CA. Email: \texttt{hdhillon@usc.edu}.}
\thanks{\IEEEauthorrefmark{3}J. G. Andrews is with the Wireless and Networking Communications Group (WNCG), The University of Texas at Austin, TX, USA. Email: \texttt{jandrews@ece.utexas.edu}.}
}

\maketitle

\begin{abstract}
Despite being ubiquitous in practice, the performance of maximal-ratio combining (MRC) in the presence of interference is not well understood. Because the interference received at each antenna originates from the same set of interferers, but partially de-correlates over the fading channel, it possesses a complex correlation structure. This work develops a realistic analytic model that accurately accounts for the interference correlation using stochastic geometry. Modeling interference by a Poisson shot noise process with independent Nakagami fading, we derive the link success probability for dual-branch {\it interference-aware} MRC. Using this result, we show that the common assumption that all receive antennas experience equal interference power underestimates the true performance, although this gap rapidly decays with increasing the Nakagami parameter $\mi$ of the interfering links. In contrast, ignoring interference correlation leads to a highly optimistic performance estimate for MRC, especially for large $\mi$. In the low outage probability regime, our success probability expression can be considerably simplified. Observations following from the analysis include: (i) for small path loss exponents, MRC and minimum mean square error combining exhibit similar performance, and (ii) the gains of MRC over selection combining are smaller in the interference-limited case than in the well-studied noise-limited case.
\end{abstract}


\IEEEpeerreviewmaketitle

\section{Introduction}\label{sec:introduction}

Diversity combining techniques are commonly used in modern wireless multi-antenna consumer devices such as smartphones, laptops and WiFi routers, to improve link reliability and energy efficiency. One of the most popular choices is maximal-ratio combining (MRC), which is known to achieve optimal performance in the absence of (multi-user) interference\cite{brennan03,simon05,goldsmith05}. In the interference-free case, MRC maximizes the post-combiner signal-to-noise ratio (SNR) by weighting the signals received at the different antennas (or equivalently, branches) according to the respective per-antenna SNRs, followed by the coherent summation of the weighted signals. Like other diversity combining schemes, MRC suffers substantial performance losses when practical non-idealities such as average reception-quality imbalance\cite{halpern77} and fading correlation\cite{aalo95} are taken into account. These performance losses are amplified further by interference, which has become a key issue with the denser usage of wireless devices; taking place particularly in non-licensed spectrum due both to offloading of cellular traffic\cite{andrews_femto} and the relentless increase of wireless consumer devices\cite{cisco13}. The main reason behind these losses is that the resulting interference is usually not equally strong across antennas because of uncorrelated or slightly correlated fading on the interferer to per-antenna links, thereby leading to additional reception-quality imbalance across the branches\cite{cui04}. Furthermore, this imbalance typically varies unpredictably fast and entails a complex correlation structure across antennas that depends upon various system parameters, such as the locations of the interferers and the fading gains.

Although information-theoretically suboptimal in the presence of interference, MRC is expected to remain a widespread diversity combining technique in the near future due to its maturity and low implementation costs compared to other competing techniques, e.g., interference-canceling combining schemes, which usually require a higher channel estimation effort. This motivates the study of the performance of MRC under a more realistic channel and interference model, which is the main focus of this paper.

\subsection{Related Work and Motivation}

The impact of interference on the performance of MRC was first studied assuming \textit{deterministic} interference power at all branches for both the equal as well as the unequal strength case\cite{cui04,cui99,aalo00}. Using the notion of outage probability, these works demonstrated that interference may severely degrade the expected performance depending on the number of interferers and their strength, especially for the case of unequal strengths. In a broader sense, the outage probability expressions derived in these works may be seen as \textit{conditional} on the interference statistics. Therefore, to evaluate the overall performance, one needs to average over the interference, which is challenging because interference depends upon various system parameters and often appears {\em random} to the receiver.

Recently, tools from stochastic geometry\cite{stoyan95} has been proposed for addressing this and other closely related challenges\cite{baccelli09a,baccelli09b,HaenggiBook,weber10,andrews11,tanbourgi13_3}. Using these tools, the performance of MRC in the presence of interference, modeled as a Poisson shot noise field, was studied in several works, mainly under two simplified interference correlation models: for instance, in\cite{sheng10,rajan10} the interference power was assumed statistically independent across the antennas, although it is correlated as the interference terms at the different antennas originate from the same source of randomness, i.e., from the same set of transmitters. This type of correlation is often neglected in the literature\cite{ganti09}, which results in significantly overestimating the true diversity. On the other hand,\cite{hunter08} assumed the same interference strength at all antennas, which corresponds to modeling the interference power as being fully correlated across the branches. This, in turn, underestimates the true diversity as the de-correlation effect of the channel fading is ignored. The importance of properly modeling interference correlation was highlighted in\cite{chopra11,chopra12,ganti09_1}. In\cite{chopra11,chopra12}, the interference properties measured at a multi-antenna receiver were analyzed within the continuum between complete independence and full correlation of the interference. In\cite{ganti09_1}, the second-order statistics of the interference and of outage events were characterized. This led for example to an exact performance evaluation of the simple retransmission scheme\cite{Haenggi14twc}, selection combining\cite{haenggi12_1} as well as cooperative relaying\cite{tanbourgi_13_1,crismani13}. 

Another frequently made assumption in the literature\cite{cui04,zhang07,ahn09,direnzo13}, is that the MRC combining weights do not dependent on the interference-plus-noise power experienced at each antenna, i.e., they are proportional only to the fading gains of the desired link. Such an MRC model may be seen as \emph{interference-blind} and is suboptimal when the interference-plus-noise power varies across antennas. In slight contrast, the MRC combining weights in\cite{chopra11,chopra13} were assumed to be additionally inversely proportional to the interferer density corresponding to the interference field seen by each antenna. Since the interferer density is proportional to the mean interference power\cite{HaenggiBook}, this form of MRC essentially performs an adaptation to the long-term effects of the interference. The authors showed that such a long-term adaptation yields some improvements when interference is correlated across antennas.

When the {\it current} per-antenna interference-plus-noise powers in one transmission period are known to the receiver, e.g., through estimation within the channel training period\cite{benedict67,pauluzzi00}, they can be taken into account when computing the MRC weights; thereby following the MRC approach of \cite{brennan03}. In\cite{tanbourgi13_2}, and in contrast to all previous works, the performance under spatial interference correlation of such an \emph{interference-aware} MRC receiver model was recently analyzed assuming Rayleigh fading channels and absence of receiver noise. For the practical dual-branch case, the exact distribution of the post-combiner signal-to-interference-plus-noise ratio ($\sinr$) was derived, while bounds were proposed for the case of more than two branches. 

\subsection{Contributions and Outcomes}
In this work, we extend the findings obtained in\cite{tanbourgi13_2} for interference-aware MRC by considering Nakagami fading and receiver noise, and discuss related design aspects with emphasis on the effect of spatial interference correlation. Similar to\cite{tanbourgi13_2}, we assume an {\it isotropic} interference model\cite{chopra12,chopra13}, i.e., each antenna sees interference from the same set of interferers, which results in interference correlation across antennas. Our main contributions and insights are summarized below.

\textit{Success probability for dual-branch MRC:} The main result of this paper is Theorem~\ref{thm:cov_prob} in Section~\ref{sec:cov_prob}, which gives an analytical expression for the exact success probability (1-outage probability) for a dual-branch MRC receiver under spatially-correlated interference, receiver noise and independent Nakagami fading. Importantly, the Nakagami fading parameter does not have to be identical for the desired and the interfering links, whereas the parameter for the desired links is restricted to integers. We show how previous results from the literature are special cases of Theorem~\ref{thm:cov_prob}. For the low outage probability regime, we derive a tractable closed-form expression for the main result later in Section~\ref{sec:asym}.

\textit{Comparison with simpler correlation models:} In Section~\ref{sec:simple_models}, we use the main result to study the accuracy loss associated with simpler correlation models frequently used due to their analytical tractability. It is shown that ignoring interference correlation across the branches results in a considerably optimistic performance characterization of MRC, particularly for large Nakagami fading parameters (small channel variability). The picture changes when assuming an identical interference level across the branches; here, the available diversity is underestimated, which yields a slightly pessimistic performance characterization. The resulting success probability gap, however, rapidly decreases with the Nakagami fading parameter of the interfering links and becomes no greater than about $10\%$ depending on the path loss exponent. This intuitive trend eventually yields an asymptotic equivalence between the full-correlation and the exact model, which is mathematically established in Section~\ref{sec:simple_models}. One important insight is that the simpler full-correlation model can be used whenever the interfering links undergo a strong path loss and/or poor scattering.

\begin{figure}[t]
	\centering
	\includegraphics[width=0.420\textwidth]{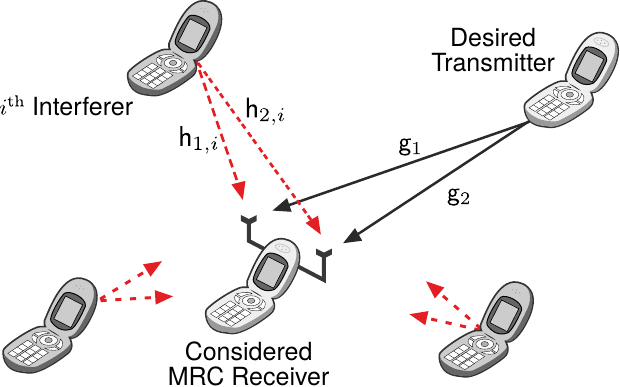}
	\caption{Illustration of the underlying scenario for the example $N=2$. The considered dual-antenna receiver is located at the origin. The desired transmitter is located $d$ meters away. The considered receiver experiences interference from surrounding interferers. 
	} \label{fig:illustration}
\end{figure}

\textit{Efficient method for semi-numerical evaluation of the result:} In Section~\ref{sec:diff}, we propose and discuss a methodology for efficient and robust semi-numerical evaluation of the result of Theorem~\ref{thm:cov_prob}. We mainly make use of Fa\`{a} di Bruno's formula, followed by a method for numerical differentiation based on Chebyshev polynomial approximation. Although immaterial to the theoretical framework, the ideas presented in this section are helpful for applying and reproducing our theoretical results using numerical software.

\textit{Comparison with other diversity combining techniques:} Using the main result for the dual-branch case, we compare the performance of MRC to other widely-known diversity combining schemes under the influence of spatial interference correlation in Section~\ref{sec:other_div_com}. We find that minimum mean square error (MMSE) combining, which does not treat interference as white noise, yields a linear diversity-gain increase with the path loss exponent compared to MRC. For small path loss exponents, there is almost no benefit from estimating and rejecting interference using MMSE as MRC, although sub-optimal, achieves almost the same diversity gain. The benefit of MRC over selection combining (SC) in terms of diversity gain is in general smaller than in the interference-free case, and monotonically decreases with the path loss exponent. For typical path loss exponents, the performance of MRC is about $1$ dB higher than for SC. Interestingly, when the path loss exponent tends to two, the gain of MRC over SC becomes equal to the corresponding value for the interference-free case.

{\bf Notation:} We use sans-serif-style letters ($\mathsf{z}$) and serif-style letters ($z$) for denoting random variables and their realizations or variables, respectively. We define $(z)^{+}\mathdef\max\{0,z\}$.

\section{System Model}\label{sec:notation}
We consider an $N$-antenna receiver communicating with a desired transmitter at an arbitrary distance $d$.\footnote{Although the main result captures only the dual-antenna case, it will be useful in the later discussions to generalize the model to $N$ antennas.} The transmitted signal received at the $N$ antennas is corrupted by noise and interference caused by other transmitters. The locations $\{\mathsf{x}_{i}\}_{i=0}^{\infty}$ of these interfering transmitters are modeled by a stationary planar Poisson point process (PPP) $\Phi\mathdef\{\mathsf{x}_{i}\}_{i=0}^{\infty}\subset\mathbb{R}^2$ of density $\lambda$. The PPP model is widely-accepted for studying multiple kinds of networks, see for instance\cite{baccelli09b,andrews11,blas12}. 
More complex interference geometries, e.g., with carrier-sensing at the nodes, can be incorporated with acceptable effort using Poisson-like models, cf.\cite{baccelli09b,hunter10,tanbourgi12}. Such modifications are beyond the scope of this contribution.

\newcounter{mycounter}
\begin{figure*}[!t]
\normalsize
\setcounter{mycounter}{\value{equation}}
\setcounter{equation}{3}
\begin{IEEEeqnarray}{rCl}
\cp&=&\sum\limits_{k=0}^{\md-1}\frac{(-1)^{k+\md}}{k!\,\Gamma(\md)}\int_{0}^{\infty}\frac{\partial^{k}\partial^{\md}}{z\,\partial s^{k}\partial t^{\md}}\left[\exp\left(-\frac{(T-z)^{+}s\md}{\snr}-\frac{zt\md}{\snr}-\pi\lambda\mathcal{A}(z,s,t)\right)\right]_{\substack{s=1\\ t=1}}\mathrm dz\label{eq:pc_theorem}\IEEEeqnarraynumspace
\end{IEEEeqnarray}
\hrulefill
\begin{subnumcases}{\label{eq:cal_A}\mathcal{A}(z,s,t) =} 
s^{2/\alpha}(T-z)^{2/\alpha}\,d^{2}\,\Gamma(1-2/\alpha)\left(\tfrac{\md}{\mi}\right)^{2/\alpha}\Gamma(2/\alpha+2\mi)\notag\\
\qquad\quad\times\,{}_{2}\mathbf{F}_{1}\left(-2/\alpha,\mi,2\mi,1-\frac{zt}{(T-z)s}\right),\quad 0\leq z< T\\
(zt)^{2/\alpha}\,d^{2}\,\Gamma(1-2/\alpha)\left(\tfrac{\md}{\mi}\right)^{2/\alpha}\frac{\Gamma(2/\alpha+\mi)}{\Gamma(\mi)},\quad z\geq T
\end{subnumcases}
\hrulefill
\begin{IEEEeqnarray}{rCl}
		\cp^{\alpha=4,m=1}&=&-\int_{0}^{\infty}z^{-1}\exp\left(-\frac{(T-z)^{+}}{\snr}\right)\frac{\partial}{\partial t}\left[\exp\left(-\frac{zt}{\snr}-\frac{\lambda\pi^2}{2}\frac{\left((T-z)^{+}\right)^{3/2}-(zt)^{3/2}}{(T-z)^{+}-zt}\right)\right]_{t=1}\,\mathrm dz\IEEEeqnarraynumspace\label{eq:special_case1}
	\end{IEEEeqnarray}
\setcounter{equation}{\value{mycounter}}
\hrulefill
\vspace*{3pt}
\end{figure*}

Due to the stationarity of $\Phi$ the interference statistics are location-invariant\cite{stoyan95}. Thus, we can place the considered receiver in the origin $o\in\mathbb{R}^2$ without loss of generality. The path loss between a given transmitter at $x\in\mathbb{R}^2$ and the considered receiver is given by $\|x\|^{-\alpha}$, where $\alpha>2$ is the path loss exponent. We denote by $\mathsf{g}_{n}$ the channel fading (power) gain between the desired transmitter and the $n\th$ antenna of the considered receiver. Similarly, the set of channel fading gains of the interfering channels to the $n\th$ antenna is defined as $\mathbf{h}_{n}\mathdef\{\mathsf{h}_{n,i}\}_{i=0}^{\infty}$, where $\mathsf{h}_ {n,i}$ denotes the fading gain of the channel between the $i\th$ interferer to the $n\th$ antenna of the considered receiver. We consider independent Nakagami fading across all channels, which corresponds to assuming that all fading gains independently follow a Gamma distribution having probability density function
\begin{align}
f_{\mathsf{y}}(y) = \frac{m^m  y^{m-1}}{\Gamma(m)} \exp \left(-m y \right),\quad y\geq0,
\end{align}
with shape $m$ and scale $1/m$, where $m$ is the Nakagami fading parameter\cite{goldsmith05}. To preserve generality, we allow for non-identical fading between the desired and the interfering links, i.e., desired and interference signals undergo Nakagami fading with possibly unequal Nakagami parameter. In what follows, the $\mathsf{g}_{n}$ are associated with Nakagami parameter $\md$, while the $\mathsf{h}_{n,i}$ are associated with Nakagami parameter $\mi$. Importantly, we require $\md$ to be integer-valued. The corresponding tail probability of $\mathsf{g}_{n}$ (similarly, $\mathsf{h}_{n,i}$) is given by $\mathbb{P}(\mathsf{g}_{n}>g)=Q(\md,\md g)$ for $n=1,\ldots,N$, where $Q(a,x)\mathdef\Gamma(a,x)/\Gamma(a)$ is the regularized upper incomplete Gamma function \cite{olver10}. It is easy to check that $\mathbb{E}[\mathsf{g}_{n}] = 1$, and $\mathsf{g}_{n} \rightarrow 1$ almost surely as $\md \rightarrow \infty$. The same holds for $\mathsf{h}_{n,i}$ for all $n=1,\ldots,N$ and $i\in\mathbb{N}$. Possible extensions toward general fading distributions can be incorporated in the model, e.g., using ideas from \cite{keeler13,dhillon13}. We assume the same fixed transmit power for all nodes and a slotted medium access with a slot duration smaller than or equal to the channel coherence time, and leave possible extensions for future work. Fig.~\ref{fig:illustration} illustrates the considered scenario.

\begin{table}[!t]
	\renewcommand{\arraystretch}{1.3}
	\caption{General notation used throughout this work}
	\label{tab:notation}
	\centering
	\small
	\begin{tabular}{c|p{6.8cm}}
		\hline
		\bfseries{Notation} & \hspace{2.4cm}\bfseries{Description}\\
		\hline
		$N$	& Number of receive antennas (branches)\\
		\hline
		$d$				& Distance between considered receiver and desired transmitter\\
		\hline
		$\alpha$			& Path loss exponent\\
		\hline
		$\mathsf{g}_{n}$			& Power fading gain between desired transmitter and $n\th$ antenna of the considered receiver\\
		\hline
		$\mathsf{h}_{n,i};\mathbf{h}_{n}$	& Power fading gain between the $i\th$ interferer and $n\th$ antenna of the considered receiver; set $\{\mathsf{h}_{n,i}\}_{i=1}^{\infty}$ of all interferer channel gains to the $n\th$ antenna of the considered receiver\\
		\hline
		$\md;\mi$				& Nakagami fading parameter on the desired links; and on the interfering links\\
		\hline
		$\Phi;\lambda$ 		& Interferer locations modeled as PPP; spatial density of interferers\\
		\hline
		$\mathsf{I}_{n}$ & Current interference power at $n\th$ antenna (branch)\\
		\hline
		$\snr$ & Average SNR at the considered receiver\\
		\hline
		$\sinrmrc$ &	Post-combiner $\sinr$ for MRC\\
		\hline
		$T$		&		$\sinr$ threshold\\
		\hline
		$\cp$	& Success probability for an MRC receiver\\
		\hline
	\end{tabular}
\end{table}

We assume that the receiver is interference-aware, i.e., it can not only perfectly estimate the instantaneous fading gain of the desired link but also the current interference-plus-noise power within one slot. By\cite{brennan03}, the MRC weight in the $n\th$ branch is proportional to the fading amplitude gain of the desired link and inversely proportional to the current interference-plus-noise power at the $n\th$ antenna, see Appendix~\ref{sec:mrc_model} for details. The post-combiner $\sinr$ for MRC then takes the form
\begin{IEEEeqnarray}{rCl}
	\sinrmrc\mathdef\frac{\mathsf{g}_{1}}{\mathsf{I}_{1}+\snr^{-1}}+\ldots+\frac{\mathsf{g}_{N}}{\mathsf{I}_{N}+\snr^{-1}},\label{eq:sir_general}\IEEEeqnarraynumspace
\end{IEEEeqnarray}
where $\mathsf{I}_{n}\mathdef d^{\alpha}\sum_{\mathsf{x}_{i}\in\Phi}\mathsf{h}_{n,i}\|\mathsf{x}_{i}\|^{-\alpha}$ is the interference power experienced at the $n\th$ antenna normalized by $d^{-\alpha}$ and $\snr$ is the average signal-to-noise ratio. $\mathsf{I}_{n}$ is understood as the instantaneous interference power averaged over the interferer symbols within one transmission slot, and hence corresponds to the current variance of the aggregate interference signal at the $n\th$ antenna, see Appendix~\ref{sec:mrc_model} for details. Due to the slotted medium access, we can assume that $\mathsf{I}_{n}$ remains constant for the duration of one slot. It can be shown that $\mathsf{I}_{n}<\infty$ almost surely for all $n\in[1,\ldots,N]$ when $\alpha>2$\cite{HaenggiBook}. Note that, although the fading gains $\mathbf{h}_{1},\ldots,\mathbf{h}_{N}$ are independently distributed, the $\mathsf{I}_{1},\ldots,\mathsf{I}_{N}$ and hence the individual $\sinr$s on different branches are correlated since the interference terms originate from the same set of interferers, i.e., from the point process $\Phi$. The distribution of \eqref{eq:sir_general} can, in general, be obtained using the joint density of the interference amplitudes derived in\cite{chopra12} for the case of isotropic interference, i.e., averaging the conditional $\sinr$ distribution over the interference statistics. However, this approach is analytically involved since (i) the joint density cannot be given in closed-form and (ii) the sum of non-identical gamma random variables must be considered. Table~\ref{tab:notation} summarizes the notation used in this work.

\section{Success Probability of Dual-Branch MRC}\label{sec:cov_prob}

In this section, the performance of MRC receivers under the setting described in Section~\ref{sec:notation} is studied. We use the success probability as the performance metric, which is defined as
\begin{IEEEeqnarray}{rCl}
	\cp\mathdef\mathbb{P}\left(\sinrmrc\geq T\right)
\end{IEEEeqnarray}
for a modulation- and coding-specific $\sinr$-threshold $T>0$. The $\cp$ can be seen as the complementary cumulative distribution function of the $\sinrmrc$ or as 1-outage probability.

The number of antennas mounted on practical wireless devices typically remains small due to space limitations and complexity constraints, e.g., smartphones, WiFi routers, thereby often not exceeding $N=2$ antennas. For this special case, the following key result characterizes the resulting performance in terms of success probability.

\begin{theorem}[Success probability of dual-branch MRC] \label{thm:cov_prob} The success probability for dual-branch MRC ($N=2$) under the described setting is given by \eqref{eq:pc_theorem} 
at the top of the page.
\end{theorem}

\begin{IEEEproof}
	See Appendix~\ref{proof:cov_prob}.
\end{IEEEproof}
\setcounter{equation}{5}
The function ${}_{2}\mathbf{F}_{1}(a,b,c;z)\mathdef{}_{2}{F}_{1}(a,b,c;z)/\Gamma(c)$ is known as the \textit{regularized} Gaussian hypergeometric function\cite{olver10} and is implemented in most numerical software programs. A method for efficient and robust semi-numerical evaluation of the success probability result of Theorem~\ref{thm:cov_prob} is presented and discussed in Section~\ref{sec:diff}.

\begin{remark}
	The integral in \eqref{eq:pc_theorem} over $[0,\infty)$ can be split into two integrals with limits $[0,T)$ and $[T,\infty)$ to get rid of the $(\cdot)^{+}$ function and to exploit the fact that the integrand of the upper integral becomes zero for all $s$-derivatives.
\end{remark}

Making use of the functional relation $\hyperg(-1/2,1,2,z)=\tfrac{2}{3z}\left(1-(1-z)^{3/2}\right)$, the result in Theorem~\ref{thm:cov_prob} can be further simplified in the case of Rayleigh fading and a path loss exponent $\alpha=4$. 
\begin{corollary}[Special case: $\alpha=4$, Rayleigh fading links] When $\md=\mi=m=1$ (Rayleigh fading) and $\alpha=4$, the success probability under the described setting for dual-branch MRC ($N=2$) reduces to \eqref{eq:special_case1} at the top of the page.
\end{corollary}\setcounter{equation}{6}
Similar simplifications that express \eqref{eq:pc_theorem} through elementary functions can be obtained by invoking functional identities of the Gaussian hypergeometric function for suitable $\alpha$ and $\mi$\cite{abramowitz64,olver10}.

\begin{remark} Letting $\snr\to\infty$ in \eqref{eq:special_case1} and differentiating with respect to $t$, we recover the result from \cite{tanbourgi13_2}. 
\end{remark}

\begin{figure}[t]
  \centering
  \includegraphics[width=.48\textwidth]{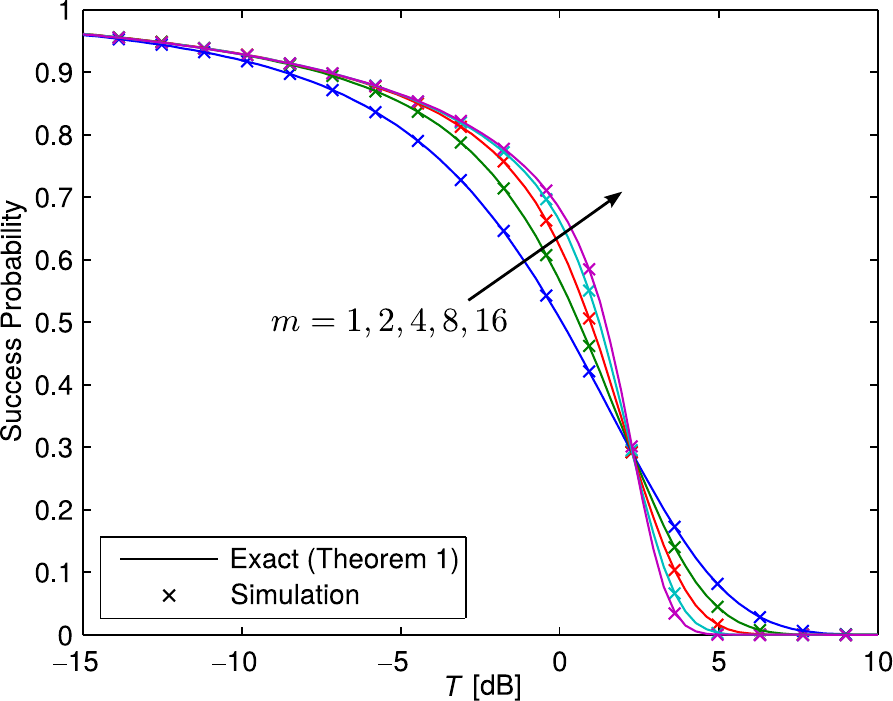}  
  \caption{$\cp$ vs. $T$ for different $\md=\mi=m$ (identical Nakagami fading). Parameters are: $\lambda=10^{-3}$, $\alpha=4$, $d=10$ $\snr=0$ dB. Marks represent simulation results.}\label{fig:pc}
\end{figure}

Figure~\ref{fig:pc} shows the success probability $\cp$ over $T$ for different $\md=\mi=m$ (identical Nakagami fading). It can be seen that the result from Theorem~\ref{thm:cov_prob} perfectly matches the simulation results. Furthermore, increasing the Nakagami fading parameter has two effects on $\cp$: for not too small values of $\cp$, decreasing channel variability ($m\uparrow$) improves transmission reliability, whereas for (non-practical) small values of $\cp$ this trend is reversed. Interestingly, all curves seem to intersect at one unique point (in this example around $T=2.3$ dB).

From the general result of Theorem~\ref{thm:cov_prob}, one can derive the success probability under pure interference-limited and pure noise-limited performance.  

\begin{corollary}[Interference vs. noise]
	The success probability $\lim_{\snr\to\infty}\cp$ in the interference-limited regime is obtained by letting $\snr\to\infty$ in \eqref{eq:pc_theorem}. Similarly, the success probability $\lim_{\lambda\to0}\cp$ in the noise-limited case can be recovered by letting $\lambda\to0$ in \eqref{eq:pc_theorem}, yielding $\cp=Q(2\md,\md T/\snr)$.
\end{corollary}
\begin{IEEEproof}
	By the dominated convergence theorem, we can interchange limit and integration in both cases. For the noise-limited case, we further note that
	\begin{IEEEeqnarray}{rCl}
		&&\sum\limits_{k=0}^{\md-1}\hspace{-.08cm}\frac{(-1)^{k+\md}}{k!\,\Gamma(\md)}\hspace{-.05cm}\int_{0}^{\infty}\hspace{-.3cm}\frac{\partial^{k}\partial^{\md}}{z\,\partial s^{k}\partial t^{\md}}\hspace{-.1cm}\left[\exp\hspace{-.05cm}\left(-\frac{s\psi_{1}}{\snr}-\frac{t\psi_{2}}{\snr}\right)\right]_{\substack{s=1\\ t=1}}\hspace{-.15cm}\mathrm dz\IEEEnonumber\\
		&&=\int_{0}^{\infty}\hspace{-.18cm}\left(\frac{\md}{\snr}\right)^{\hspace{-.07cm}\md}\hspace{-.07cm}\frac{z^{\md-1}e^{-\frac{z\md}{\snr}}}{\Gamma(\md)}\,Q\hspace{-.1cm}\left(\md,\frac{\md}{\snr}(T-z)^{+}\right)\,\mathrm dz\IEEEnonumber\\
		&&=\mathbb{E}_{\mathsf{g}_{2}\snr}\big[\mathbb{P}_{\mathsf{g}_{1}\snr}\left(\mathsf{g}_{1}\snr+\mathsf{g}_{2}\snr\geq T\left\lvert\right.\mathsf{g}_{2}\snr\right)\big]\IEEEnonumber\\
		&&=Q\left(2\md,\tfrac{\md T}{\snr}\right)
	\end{IEEEeqnarray}
which concludes the proof.
\end{IEEEproof}

Another special case one may think of is when the channel variability becomes very small, i.e., $1/\md,1/\mi\to0$, eventually leading to the pure path loss model. However, taking the limit $\md,\mi\to\infty$ in \eqref{eq:pc_theorem} looks quite difficult. 

\begin{remark}[Success Probability as $\md,\mi\to\infty$] 
	Since $\mathsf{g}_{n} \rightarrow 1$ and $\mathsf{h}_{n} \rightarrow 1$ as $\md,\mi \rightarrow \infty$, the $\sinrmrc$ of a $N$-branch receiver becomes $\tfrac{N}{\snr^{-1}+ \mathsf{I}}$, with $\mathsf{I}=d^{\alpha}\sum_{\mathsf{x}_{i}\in\Phi}\|\mathsf{x}_{i}\|^{-\alpha}$, which is the same as the $\sinr$ of a single-branch receiver with $N$-fold received power increase. The corresponding $\cp$ can be characterized, e.g., by Laplace inversion\cite{baccelli09a} or by the dominant-interferer bounding technique\cite{weber10}. For the case of $\alpha=4$, a closed-form solution can be found in\cite{HaenggiBook}.
\end{remark}

\section{Comparison with Simpler Correlation Models}\label{sec:simple_models}
For analytical tractability, it is frequently assumed in the literature that the interference power across different branches is either equally-strong or statistically independent. Certainly, such simplifications may lead to an accuracy loss as the true interference correlation structure is distorted. Using the exact model derived in Section~\ref{sec:cov_prob}, this accuracy loss is studied next.\vspace{-.2cm}

\subsection{Full-Correlation Model}
In the full-correlation model, the current interference power is assumed equally strong across the branches, i.e., $\mathsf{I}_{n}\equiv\mathsf{I}_{m}$ for $m,n\in[1,\ldots,N]$, see for instance\cite{hunter08,crismani13}. This assumption effectively ignores the additional variability in the per-branch $\sinr$s resulting from the de-correlation effect of the fading on the interfering links. 

\begin{definition}[Full-correlation (FC) model]\label{def:fc}
	In the FC model, the interference terms $\mathsf{I}_{n}$ at the $N$ branches are assumed to be equal, i.e., $\mathsf{h}_{m,i}\equiv\mathsf{h}_{n,i}$ for all $m,n\in[1,\ldots,N]$ and $i\in\mathbb{N}$. The corresponding post-combiner $\sinr$ is $\sinrfc$.
\end{definition}

Hence, in the FC model the post-combiner $\sinr$ becomes
\begin{IEEEeqnarray}{rCl}
	\sinrfc=\frac{\sum_{n=1}^{N}\mathsf{g}_{n}}{\mathsf{I}+\snr^{-1}}.
\end{IEEEeqnarray}

The next result gives the success probability $\cpfc$ in the FC model for arbitrary $N\geq1$.

\begin{proposition}[Success probability $\cpfc$ for FC model]\label{prop:cov_prob_fc}
	The success probability for $N$-branch MRC in the FC model is
	\begin{IEEEeqnarray}{rCl}
		\cpfc&=&\hspace{-.15cm}\sum\limits_{k=0}^{N\md-1}\hspace{-.1cm}\frac{(-1)^{k}}{k!}\,\frac{\partial^k}{\partial s^{k}}\left[\exp\left(-\frac{s\md T}{\snr}-\lambda\pi d^{2}s^{2/\alpha}\right.\right.\IEEEnonumber\\
	&&\left.\left.\hspace{-.15cm}\times T^{2/\alpha}\Gamma(1-2/\alpha)\frac{\Gamma(2/\alpha+\mi)}{\Gamma(\mi)}\left(\tfrac{\md}{\mi}\right)^{2/\alpha}\right)\right]_{s=1}\hspace{-.3cm}.\IEEEeqnarraynumspace
	\end{IEEEeqnarray}
\end{proposition}
\begin{IEEEproof}
	We first note that $\sum_{n=1}^{N}\mathsf{g}_{n}$ is Gamma distributed with shape parameter $N\md$ and scale parameter $1/\md$\cite{feller71}. Applying a similar technique as in the proof of Theorem~\ref{thm:cov_prob}, we obtain
	\begin{IEEEeqnarray}{rCl}
		\cpfc&=&\mathbb{E}_{\mathsf{I}}\big[Q\left(N\md,\md T(\mathsf{I}+\snr^{-1})\right)\big]\IEEEnonumber\\
		&=&\sum\limits_{k=0}^{N\md-1}\frac{(-1)^{k}}{k!}\,\frac{\partial^{k}}{\partial s^{k}}\big[\mathcal{L}_{\mathsf{Y}}(s)\big]_{s=1},
	\end{IEEEeqnarray}
	where $\mathsf{Y}\mathdef \md T\,(\mathsf{I}+\snr^{-1})$. Finally, the Laplace transform $\mathcal{L}_{\mathsf{Y}}(s)$ is computed using the probability generating functional (PGFL) of a PPP \cite{stoyan95}.
\end{IEEEproof}

\begin{figure*}[!t]
	\centerline{\subfloat[Success Probability]{\includegraphics[width=.48\textwidth]{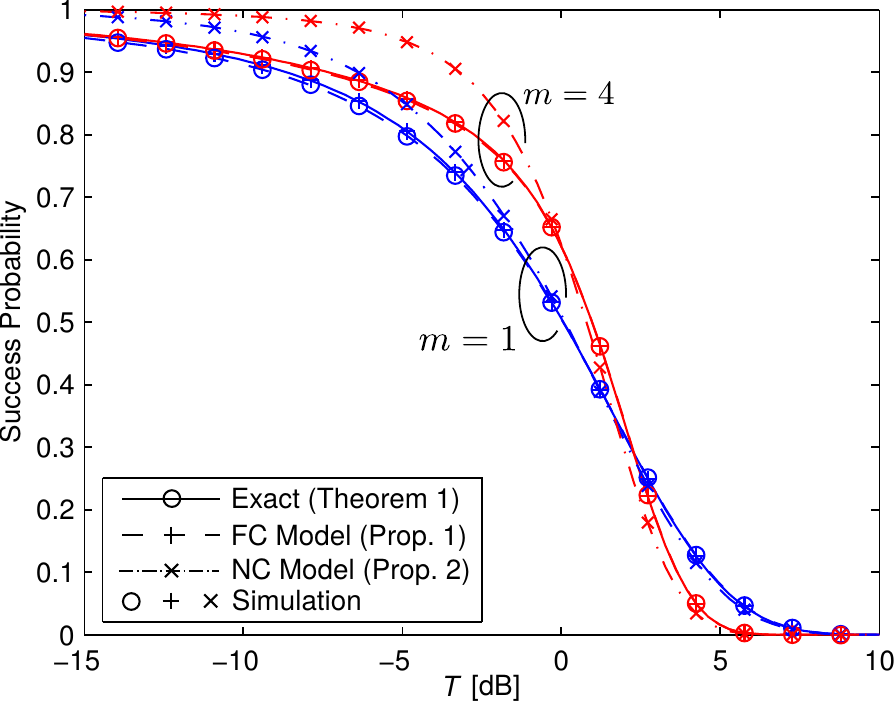}
	\label{fig:comp_exact_fc_nc}}
	\hfil
	\subfloat[FC Outage Probability Deviation]{\includegraphics[width=0.48\textwidth]{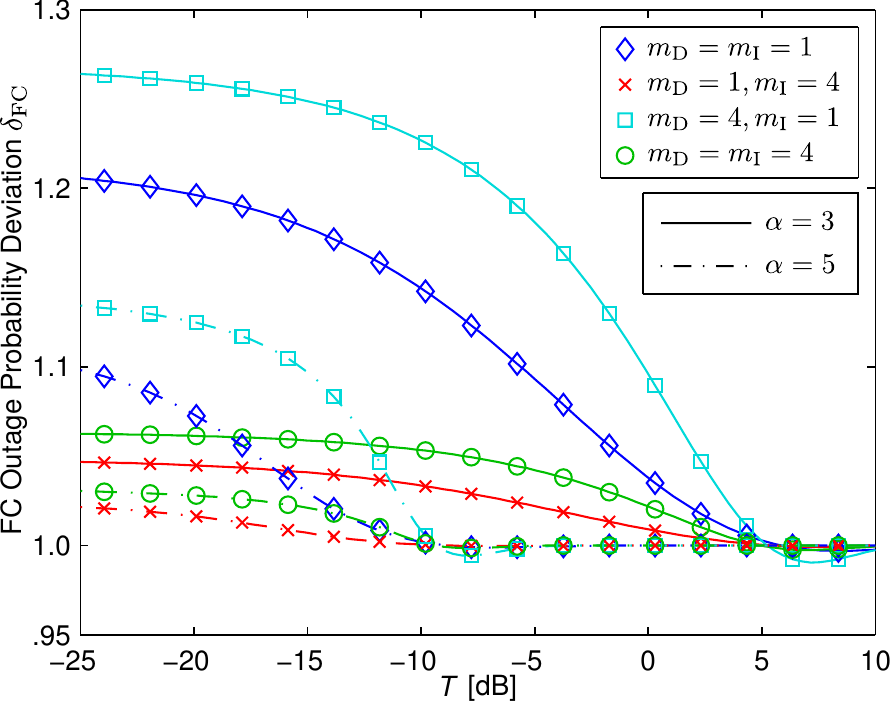}
	\label{fig:dev_mrc_fc}}}
	\caption{(a) Success probability vs. $\sinr$-threshold $T$ for different $\md=\mi=m$. Marks represent simulation results. Parameters are: $\lambda=10^{-3}$, $\alpha=4$, $d=10$, $\snr=0$ dB. (b) Outage probability deviation of FC model vs. $\sinr$-threshold $T$ for different $\md$, $\mi$, and $\alpha$. Parameters are $\lambda=10^{-3}$, $d=10$, $\snr=10(4-\alpha)$.}
	\vspace*{4pt}
\end{figure*}

\subsection{No-Correlation Model}
In contrast to modeling the interference terms $\mathsf{I}_{n}$ as being (fully) correlated, one can also assume statistical independence among them. Then, \eqref{eq:sir_general} reduces to a sum over i.i.d. random variables. Note that this no-correlation model overestimates the true diversity.

\begin{definition}[No-correlation (NC) model]\label{def:nc}
	In the NC model, the interference terms $\mathsf{I}_{n}$ at the $N$ branches are assumed to be statistically independent, i.e., $\mathbb{P}\left(\{\mathsf{I}_{n}\in A\}\cap\{\mathsf{I}_{m}\in B\}\right)=\mathbb{P}\left(\mathsf{I}_{n}\in A\right)\,\mathbb{P}\left(\mathsf{I}_{m}\in B\right)$ for all $m,n\in[1,\ldots,N]$ and all Borel sets $A,B$ on $\mathbb{R}^{+}_{0}$. The corresponding post-combiner $\sinr$ is denoted by $\sinrnc$.
\end{definition}

Note that Definition~\ref{def:nc} implies that the interference experienced at each branch originates from a distinct interferer set $\{\mathsf{x}_{i}\}_{i=0}^{\infty}$. For $N>1$, one can in general (numerically) obtain the success probability $\cpnc$ by the Laplace inversion technique for sums of independent random variables, provided the Laplace transform of the per-antenna $\sinr$ is known.

\begin{proposition}[Success probability $\cpnc$ for NC model and $N=2$]\label{prop:cov_prob_nc}
The success probability for dual-branch MRC in the NC model has the same form as in \eqref{eq:pc_theorem} of Theorem~\ref{thm:cov_prob} with $\mathcal{A}(z,s,t)$ replaced by
\begin{IEEEeqnarray}{rCl}
\mathcal{B}(z,s,t) &=& \Gamma(1-2/\alpha)\,d^{2}\,\frac{\Gamma(2/\alpha+\mi)}{\Gamma(\mi)}\,\left(\frac{\md}{\mi}\right)^{2/\alpha}\IEEEnonumber\\
&&\times\left( \left(s\,(T-z)^{+}\right)^{2/\alpha}+(zt)^{2/\alpha}\right).\IEEEeqnarraynumspace\label{eq:prob_nc}
\end{IEEEeqnarray}
\end{proposition}
\begin{IEEEproof}
	The proof is analogous to the proof of Theorem~\ref{thm:cov_prob} until step (a) in \eqref{eq:proof_thm1_step7}. Due to distinct interferer sets across the two branches, the expectation with respect to $\Phi$ in \eqref{eq:proof_thm1_step7} step (a) decomposes into the product
	\begin{IEEEeqnarray}{rCl}
		&&\mathbb{E}_{\Phi}\left[\prod\limits_{\mathsf{x}_{i}\in\Phi}\mathbb{E}_{\mathsf{h}_{1}}\Big[\exp\left(-s\psi_{1}d^{\alpha}\mathsf{h}_{1}\|\mathsf{x}_{i}\|^{-\alpha}\right)\Big]\right]\IEEEnonumber\\
&&\quad\times\mathbb{E}_{\Phi}\left[\prod\limits_{\mathsf{x}_{i}\in\Phi}\mathbb{E}_{\mathsf{h}_{2}}\Big[\exp\left(-t\psi_{2}d^{\alpha}\mathsf{h}_{2}\|\mathsf{x}_{i}\|^{-\alpha}\right)\Big]\right]\IEEEnonumber\\
		&&\hspace{.2cm}\overset{\text{(a)}}{=}\exp\left(-\lambda\pi\int_{0}^{\infty}2r\,\left(2-\mathbb{E}_{\mathsf{h}_{1}}\left[e^{-s\psi_{1}d^{\alpha}\mathsf{h}_{1}r^{-\alpha}}\right]\right.\right.\IEEEnonumber\\
&&\left.\left.\qquad\qquad\qquad\qquad\qquad-\mathbb{E}_{\mathsf{h}_{2}}\left[e^{-t\psi_{2}d^{\alpha}\mathsf{h}_{2}r^{-\alpha}}\right]\right)\,\mathrm dr\right),\IEEEeqnarraynumspace\label{eq:nc_proof}
	\end{IEEEeqnarray}
	where (a) follows from the PGFL for PPPs\cite{stoyan95}. After evaluating the integral with respect to $r$ and using the fact that $\mathbb{E}[\mathsf{h}_{n}^{2/\alpha}]=\mi^{-2/\alpha}\Gamma(2/\alpha+\mi)/\Gamma(\mi)$, \eqref{eq:nc_proof} becomes $\exp\left(-\lambda\pi\,\mathcal{B}(z,s,t)\right)$.  Substituting this back into \eqref{eq:proof_thm1_step7} step (a) proves the result.
\end{IEEEproof}

Figure~\ref{fig:comp_exact_fc_nc} compares the success probability for the exact model against the success probability for the NC and FC correlation models introduced above. The simulation results (indicated by marks) confirm our theoretical expressions. It can be seen that the NC model is considerably optimistic for practically relevant $\cp$ values. Interestingly, the gap between $\cp$ and $\cpnc$ increases with the Nakagami parameter. This is due to the fact that the de-correlation effect of the channel fading is reduced as $\mi$ increases which, in turn, increases the correlation across the per-antenna $\sinr$s. Ignoring correlation hence becomes even more inappropriate as the true diversity is strongly overestimated in this case.

In contrast, Fig.~\ref{fig:comp_exact_fc_nc} suggests that the FC model yields a closer approximate characterization of $\cp$; the gap between $\cp$ and $\cpfc$ remains fairly small over a wide range of $T$. In\cite{tanbourgi13_2} it was shown for the case $\md=\mi=1$ that the size of this gap depends on the path loss exponent $\alpha$ and ranges from $9\%$ for $\alpha=6$ to $27\%$ for $\alpha=2.5$. For larger Nakagami fading parameters the gap seems to vanish, as the $\cp$ and $\cpfc$ lines become indistinguishable already for $\md=\mi=4$. This observation motivates the following corollary.

\begin{corollary}[Asymptotic equivalence between exact and FC model]\label{col:asym_equiv}
	The exact and the FC model become asymptotically equivalent in terms of success probability as $\mi\to\infty$.
\end{corollary}
\begin{IEEEproof}
	We first consider the Laplace transform of $\mathsf{H}$ in \eqref{eq:frac_mom_H} of Appendix~\ref{proof:cov_prob} as $\mi\to\infty$. Since $\lim_{\mi\to\infty}\mathcal{L}_{\mathsf{H}}(u)=\exp\left(-u\,(s\psi_{1}+t\psi_{2})\right)$, this implies that $\mathsf{H}$ converges in distribution to a degenerative random variable with density $\delta(s\psi_{1}+t\psi_{2})$. Since $\mathsf{H}$ is uniformly integrable for all $\mi\geq1$, it then follows from \cite[Theorem~5.9]{gut05} that
\begin{IEEEeqnarray}{rCl}
	\lim_{\mi\to\infty}\mathbb{E}\left[\mathsf{H}^{2/\alpha}\right]=(s\psi_{1}+t\psi_{2})^{2/\alpha}.\label{eq:asym_tight1}\IEEEeqnarraynumspace
\end{IEEEeqnarray}
On the other hand, using the same approach as in the proof of Theorem~\ref{thm:cov_prob} until step (a) in \eqref{eq:proof_thm1_step7}, $\cpfc$ can be written as
\begin{IEEEeqnarray}{rCl}
	&&\sum\limits_{k=0}^{\md-1}\frac{(-1)^{k+\md}}{k!\,\Gamma(\md)}\int_{0}^{\infty}\frac{\partial^{k}\partial^{\md}}{z\,\partial s^{k}\partial t^{\md}}\,\Bigg[\exp\left(-\frac{s\psi_{1}}{\snr}-\frac{t\psi_{2}}{\snr}\right)\IEEEnonumber\\ 
&&\,\times\mathbb{E}_{\Phi}\bigg[\prod\limits_{\mathsf{x}_{i}\in\Phi}\hspace{-.05cm}\mathbb{E}_{\mathsf{h}}\Big[\exp\left(-(s\psi_{1}+t\psi_{2})d^{\alpha}\mathsf{h}\|\mathsf{x}_{i}\|^{-\alpha}\right)\Big]\bigg]\Bigg]_{\substack{s=1\\ t=1}}\hspace{-.25cm}\mathrm dz,\IEEEeqnarraynumspace\label{eq:proof_equiv1}
\end{IEEEeqnarray}
where we have exploited the fact that $\mathsf{h}_{m,i}\equiv\mathsf{h}_{n,i}$ for all $m,n\in[1,\ldots,N]$ and $i\in\mathbb{N}$ by Definition~\ref{def:fc}. Using the PGFL for PPPs\cite{stoyan95}, the expectation with respect to $\Phi$ in \eqref{eq:proof_equiv1} can be computed as
\begin{IEEEeqnarray}{rCl}
	&&\exp\left(-\lambda\pi\int_{0}^{\infty}2r\left(1-\mathbb{E}_{\mathsf{h}}\left[e^{-(s\psi_{1}+t\psi_{2})d^{\alpha}\mathsf{h}\|\mathsf{x}_{i}\|^{-\alpha}}\right]\right)\,\mathrm dr\right)\IEEEnonumber\\
&&\qquad=\exp\bigg(-\lambda\pi(s\psi_{1}+t\psi_{2})^{2/\alpha}d^{2}\IEEEnonumber\\
&&\qquad\qquad\qquad\quad\times\Gamma(1-2/\alpha)\frac{\Gamma(2/\alpha+\mi)}{\mi^{2/\alpha}\Gamma(\mi)}\bigg)\IEEEeqnarraynumspace
\end{IEEEeqnarray}
and shown to converge to $\exp(-\lambda\pi(s\psi_{1}+t\psi_{2})^{2/\alpha}\,d^{2}\,\Gamma(1-2/\alpha)$ as $\mi\to\infty$. Combining this observation for the FC model with the fact that after substituting \eqref{eq:asym_tight1} into \eqref{eq:poisson_mean2} the same expression is obtained for the exact model, the asymptotic equivalence of the two models follows.
\end{IEEEproof}

\begin{figure*}[!t]
\normalsize
\setcounter{mycounter}{\value{equation}}
\setcounter{equation}{16}
\begin{subnumcases}{\label{eq:diff_A}\left.\frac{\partial^{n}\mathcal{A}(z,s,t)}{\partial t^{n}}\right\lvert_{t=1} =}
(-1)^{n} z^{2/\alpha}\,d^{2}\,\Gamma(1-2/\alpha)\,\left(\frac{\md}{\mi}\right)^{2/\alpha}\frac{(-2/\alpha)_{n}(\mi)_{n}}{\Gamma(2\mi+n)}\,\Gamma(2/\alpha+2\mi)\notag\\
\qquad\qquad\quad\times{}_{2}{F}_{1}\left(-2/\alpha+n,\mi,2\mi+n,1-\frac{(T-z)s}{z}\right),\quad 0\leq z< T\\
z^{2/\alpha}\,d^{2}\,\Gamma(1-2/\alpha)\left(\frac{\md}{\mi}\right)^{2/\alpha}\frac{\Gamma(2/\alpha+\mi)}{\Gamma(\mi)}\,(2/\alpha-n+1)_{n},\quad z\geq T
\end{subnumcases}
\setcounter{equation}{\value{mycounter}}
\hrulefill
\setcounter{tmp_equation}{\value{equation}}
\setcounter{equation}{21}
\begin{IEEEeqnarray}{rCl}
	C_k\mathdef \int_{0}^{1}u^{2/\alpha-1-k}\left(1-u\right)^{k}\,\hypergeombf{-\tfrac{2}{\alpha}+\md+k}{\mi+k}{2\mi+\md+k}{\tfrac{2u-1}{u}}\,\mathrm du\label{eq:asym_cp2}
\end{IEEEeqnarray}
\setcounter{equation}{\value{tmp_equation}}
\hrulefill
\vspace*{4pt}
\end{figure*}

Corollary~\ref{col:asym_equiv} is particularly useful for justifying the use of the FC model for scenarios in which the interfering links undergo poor scattering. The remaining accuracy loss with respect to the exact model can be further studied by looking at the outage probability deviation $\delta_{\text{FC}}\mathdef(1-\cpfc)/(1-\cp)$.

Fig~\ref{fig:dev_mrc_fc} illustrates the impact of $\md$, $\mi$ and $\alpha$ on the deviation $\delta_{\text{FC}}$. In accordance with\cite{tanbourgi13_2}, the deviation decreases with $\alpha$ and/or $T$ which is due to the fact that interference power becomes effectively dominated by a few nearby interferers only; with a smaller set of interferers the interference naturally becomes more correlated. Note that the deviation $\delta_{\text{FC}}$ becomes negative for sufficiently large $T$ (practically non-relevant low $\cp$ values). This observation for the FC model is consistent with the findings in\cite{tanbourgi13_2,crismani13}. Furthermore, it can be seen how non-identical Nakagami fading affects the deviation: similar to what was observed in Fig.~\ref{fig:comp_exact_fc_nc} for the case of identical Nakagami fading, the deviation decreases with smaller variability of the fading on the interfering links, i.e., as $\mi$ increases.

Interestingly, this is not true for the fading on the desired links as the deviation increases with $\md$. This is due to the fact that for a smaller variability of fading on the desired links, the ``modeling error'' associated with the FC model becomes more salient. In this example, the additional deviation compared to the identical Nakagami case is about $5\%$ for $\alpha=5$. Hence, the FC model is inappropriate when fading variability on the desired links is smaller than on the interfering links, for instance when channel-inversion power control is used.

\section{Discussion}\label{sec:numerical}
In order to complement the theoretic work presented in the prior sections, we will discuss some related practical aspects next. First, a method for efficiently computing the result of Theorem~\ref{thm:cov_prob} is presented. Furthermore, we study the performance of dual-branch MRC in the low outage probability regime. Then, we compare the performance of MRC to other popular combining methods under a similar interference and fading setting. Finally, we also study the local throughput of dual-branch MRC receivers.

\subsection{Semi-Numerical Evaluation of Theorem~\ref{thm:cov_prob}}\label{sec:diff}
The mathematical form of \eqref{eq:pc_theorem} in Theorem~\ref{thm:cov_prob} involves two higher-order derivatives of a composite function which renders an analytical calculation of $\cp$ complicated. To compute $\cp$ for a set of parameters, one thus has to resort to numerical methods, of which several approaches exist in the literature. We next propose and discuss a methodology for efficient and robust semi-numerical evaluation of \eqref{eq:pc_theorem}.

\textbf{Fa\`{a} di Bruno's formula and Bell polynomials for analytical $t$-differentiation:} High-order derivatives of general composite functions of the form $f(g(x))$ can be evaluated using the well-known Fa\`{a} di Bruno formula, see for instance\cite{bruno1857,olver10}. Whenever the outer function $f(\cdot)$ is an exponential function (as in our case), it is useful to rewrite Fa\`{a} di Bruno's formula using the notion of Bell polynomials\cite{johnson07}
\begin{IEEEeqnarray}{rCl}
 \frac{\partial^{n}}{\partial x^{n}}f(g(x)) = f(g(x))\,B_{n}\left(g^{(1)}(x),\ldots,g^{(n)}(x)\right),\IEEEeqnarraynumspace\label{eq:faa_bell}
\end{IEEEeqnarray}
where $B_{n}\left(x_{1},\ldots,x_{n}\right)$ is the $n\th$ \textit{complete} Bell polynomial. The complete Bell polynomials can be efficiently obtained using a matrix determinant identity\cite{ivanoff58}. It remains to compute the derivatives of the inner function $g(x)$ up to order $n$. Transferred to our case, we thus need to compute the derivatives of the exponent in \eqref{eq:pc_theorem} up to order $\md$.

\begin{corollary}[$n\th$ $t$-derivative of $\mathcal{A}(z,s,t)$]
	The $n\th$ $t$-derivative of $\mathcal{A}(z,s,t)$ evaluated at $t=1$ is given in \eqref{eq:diff_A} at the top of the next page, where $(a)_{n}\mathdef\Gamma(a+n)/\Gamma(a)$ is the Pochhammer symbol\cite{olver10}.
\end{corollary}
Using the approach described above, the $t$-differentiation is computed analytically, i.e., without numerical difference methods. For the subsequent $s$-differentiation, however, Fa\`{a} di Bruno's formula may not be the best choice since the outer function is no longer an exponential function and the derivatives of the inner function are difficult to obtain. We therefore propose a different approach for the $s$-differentiation.  

\textbf{Chebyshev interpolation method for numerical $s$-differentiation:} Before explaining this differentiation technique, we first note that the $\partial^{k}/\partial s^{k}$ operator in \eqref{eq:pc_theorem} can be moved outside the $z$-integration according to Leibniz's integration rule for improper integrals\cite{olver10}. This step comes with the advantage of first numerically computing the integral without caring about how to perform the $s$-differentiation. Interpreting the integration result as a function of $s$, say $V(s)$, we then propose to approximate this function using the Chebyshev interpolation method in an interval $[a,b]$, yielding the approximation\cite{press07}
\setcounter{equation}{17}
\begin{IEEEeqnarray}{rCl}
	V(s)\approx\tilde V(s)\mathdef-\frac{c_{0}}{2}+\sum\limits_{i=0}^{p-1}c_{\ell}\,T_{\ell}\left(\tfrac{s-(a+b)/2}{(b-a)/2}\right),\IEEEeqnarraynumspace\label{eq:cheby_ap}
\end{IEEEeqnarray}
where $s\in[a,b]$, $T_{\ell}(x)\mathdef\cos(\ell\arccos x)$ is the $\ell\th$ Chebyshev polynomial of the first kind, $p$ is the number of sampling points, and \begin{IEEEeqnarray}{rCl}
	c_{\ell}&=&\frac{2}{p}\sum\limits_{i=0}^{p-1}V\left(\tfrac{1}{2}(b-a)\cos\left[\tfrac{\pi}{p}(i+1/2)\right]+\tfrac{1}{2}(a+b)\right)\IEEEnonumber\\
	&&\qquad\times\cos\left[\tfrac{\ell\pi}{p}(i+1/2)\right]\label{eq:cheby_nodes}
\end{IEEEeqnarray}
is the $\ell\th$ Chebyshev node. Differentiating $\tilde V(s)$ in \eqref{eq:cheby_ap} instead of $V(s)$ at the point $s=1$, we then obtain
\begin{IEEEeqnarray}{rCl}
	\left.\frac{\partial^{k}V(s)}{\partial s^{k}}\right\lvert_{s=1}\hspace{-.15cm}&\approx&\left.\frac{\partial^{k}\tilde V(s)}{\partial s^{k}}\right\lvert_{s=1}\IEEEnonumber\\
	&=&\sum\limits_{\ell=0}^{p-1}c_{\ell}\,\frac{\partial^{k}}{\partial s^{k}}\left[T_{\ell}\left(\tfrac{s-(a+b)/2}{(b-a)/2}\right)\right]_{s=1}\IEEEnonumber\\
	&\overset{\text{(a)}}{=}&\left(\frac{2}{b-a}\right)^{k}\,\sum\limits_{\ell=k}^{p-1}c_{\ell}\,T_{\ell}^{(k)}\left(\tfrac{1-(a+b)/2}{(b-a)/2}\right),\IEEEeqnarraynumspace\label{eq:cheby_der}
\end{IEEEeqnarray}
 where (a) follows from the fact that $\partial^{k}T_{\ell}(s)/\partial s^{k}=0$ when $\ell<k$ for all $s$. It is well-known that the Chebyshev approximation has the smallest maximum error among all polynomial approximations. This is due to the fact that end-points are effectively avoided through projecting the function's domain onto the angular interval $[0,\pi]$; thereby achieving exponential convergence as $p$ increases\cite{press07}.
 
 A step-by-step overview of the proposed methodology for evaluating \eqref{eq:pc_theorem} is depicted in Fig.~\ref{alg:cp}. All numerical results and figures in this work were obtained using this methodology.\vspace{.1cm}
 
{\it Some comments regarding the numerical recipe in Fig.~\ref{alg:cp}:}
\begin{itemize}
	\item Line 2: We exploit the fact that the higher-order $s$-differentiation can be moved outside the integral. This is especially useful because the $z$-integration can be efficiently computed using powerful build-in numerical integration tools with maximum-error criterion.
	\item Line 6: We used $p=\md+5$ throughout this work, which was found to yield a good balance between complexity and accuracy. Furthermore, we set $a=.8$ and $b=1.2$. 
	\item Lines 7--9: This ``for''-loop is the most time-consuming task and should be parallelized whenever allowed by the hardware and numerical software.
	\item Line 18: When $\snr<\infty$, the linear combination of $\snr$-related term and $\mathcal{A}(z,s,t)$ in the exponent of \eqref{eq:pc_theorem} must be differentiated at $t=1$. The former has first-order derivative $z\md/\snr$ and higher-order derivatives equal to zero.    
\end{itemize}

\begin{figure}[t]
\small
\begin{spacing}{1.1}

\begin{boxedalgorithmic}
\Procedure {Evaluation of \eqref{eq:pc_theorem}}{}
\State{$w_{0},\ldots,w_{\md-1}\gets s\text{-\textsc{Diff}}(\md)$}
\State{$\cp=\sum_{k=0}^{\md-1}(-1)^{k+\md}\frac{w_k}{k!\Gamma(\md)}$}
\EndProcedure
\Statex
\Function{$s$-Diff}{$\md$}\Comment{$s$-derivatives up to order $\md-1$}
\State{$s\gets[a,\ldots,b]$}\Comment{Chebyshev points, $0<a<1<b$}
\ForP{$\ell\gets0,p-1$}
	\State $V[\ell]\gets \int_0^{\infty}t\text{-\textsc{Diff}}(z,s[\ell])\,\tfrac{\mathrm dz}{z}$\Comment{Values at Chebyshev points}
\EndForP
\State{$c_{1},\ldots,c_{p}\gets\eqref{eq:cheby_nodes}$}\Comment{Get all Chebyshev nodes}
\For{$k\gets0,\md-1$}	
	\State{$\partial^{k}\tilde V(s)/\partial s^{k}|_{s=1}\gets\eqref{eq:cheby_der}$}\Comment{Differentiate interpolant}
\EndFor
\EndFunction
\Statex
\Function{$t$-Diff}{$z,s$}\Comment{$\md$-th $t$-derivative for specific $z,s$}
	\State{$f(x)\gets e^{x}$} 
	\State{$g^{(1)}(1),\ldots,g^{(\md)}(1)\gets$ \eqref{eq:diff_A} }\Comment{Get inner $t$-derivatives}	\State{$\frac{\partial^{\md}}{\partial t^{\md}}f(g(t))\gets$ \eqref{eq:faa_bell}}\Comment{Invoke Fa\`{a} di Bruno's formula}
\EndFunction
\end{boxedalgorithmic}
\caption{Numerical recipe for proposed semi-numerical evaluation of \eqref{eq:pc_theorem}.}\label{alg:cp}\vspace{-.3cm}
\end{spacing}
\end{figure}

\subsection{Asymptotic Analysis of Dual-Branch MRC}\label{sec:asym}
Practical communications systems typically operate at rather small outage probabilities in order to be energy-efficient. It is therefore interesting to study the performance of MRC in the small outage probability regime, i.e., when $\cp\to1$. A second motivation for such an asymptotic analysis is that the resulting asymptotic outage probability expression often follows a fairly simple law that can be characterized in closed-form. In this regard, it would be advantageous to obtain an asymptotic expression for $\cp$ in \eqref{eq:pc_theorem} that does no longer contain an improper-integral over two higher-order derivatives. In the following, we will consider the asymptotic performance of dual-branch MRC in the absence of receiver noise. A similar though more bulky expression can be derived also for the case with receiver noise, however, with no additional insights.

\begin{figure*}[!t]
	\centerline{\subfloat[Asymptotic Outage Probability]{\includegraphics[width=0.48\textwidth]{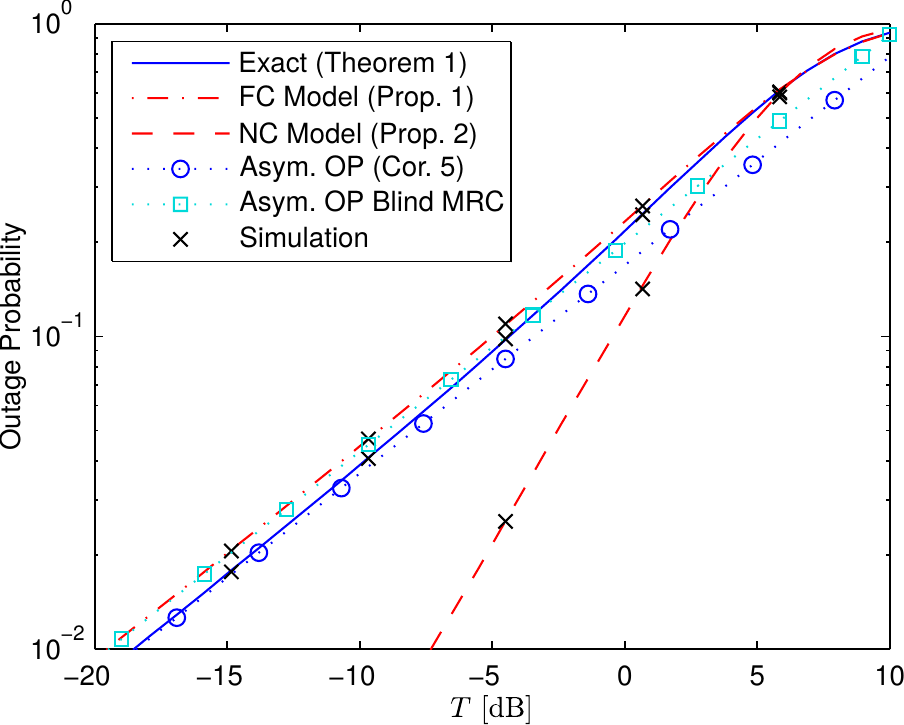}
	\label{fig:asym_ps}}
	\hfil
	\subfloat[Relative Outage Probability Reduction]{\includegraphics[width=0.49\textwidth]{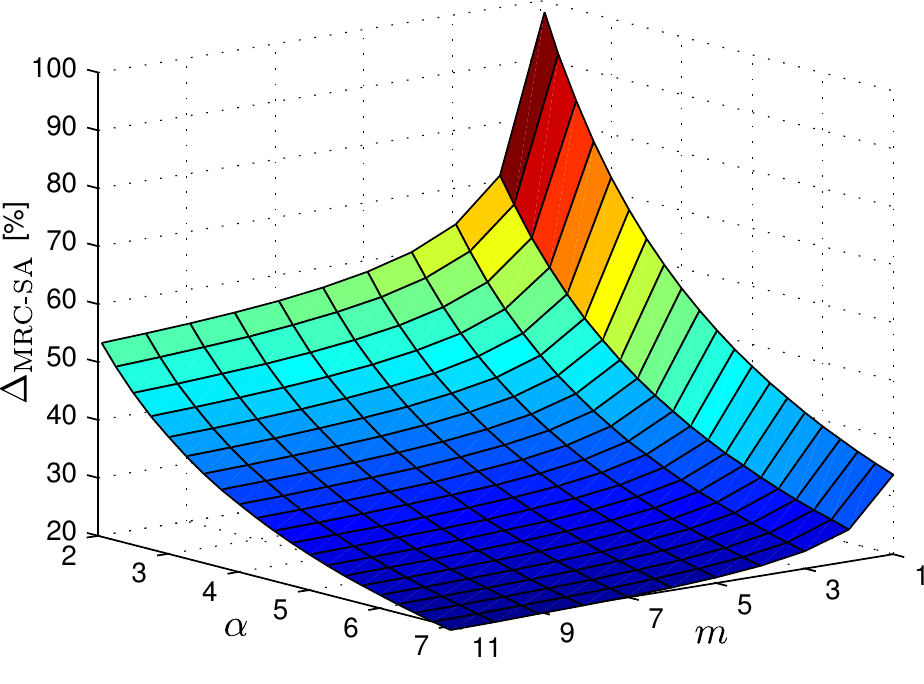}
	\label{fig:rel_gain_single_antenna}}}
	\caption{(a) Outage probability of dual-branch MRC in the low outage regime for exact, FC, and NC model. ``Blind MRC" corresponds to $1-\cp^{\text{blind}}(2)$ in \eqref{eq:chopra}. Parameters are: $\lambda=10^{-3}$, $d=10$, $\alpha=3.5$, $\md=4$, $\mi=1.5$. No receiver noise. (b) Relative outage probability reduction $\Delta_{\text{MRC-SA}}$ when switching from single-antenna to dual-branch MRC. Nakagami parameters are $\md=\mi=m$ (symmetric case). No receiver noise.}
\end{figure*}

\begin{corollary}[Asymptotic $\cp$]\label{cor:asym_cp} In the absence of noise, the success probability for dual-branch MRC under the described setting becomes
	\begin{IEEEeqnarray}{rCl}
		\cp&\hspace{-.05cm}\sim&\hspace{-.05cm}1- \kappa\,T^{2/\alpha}\,\frac{\Gamma(\md-\tfrac{2}{\alpha})\Gamma(\mi+\tfrac{2}{\alpha})}{\Gamma(\mi)\,\Gamma(\md)}\IEEEnonumber\\
		&&\hspace{-.45cm}+\tfrac{2}{\alpha}\kappa\,T^{2/\alpha}\frac{\Gamma(2\mi+\tfrac{2}{\alpha})}{B(\mi,\md)}\hspace{-.05cm}
		\sum\limits_{k=0}^{\md-1}\hspace{-.12cm}\frac{\Gamma(-\tfrac{2}{\alpha}+\md+k)\,C_k}{B(\mi,k+1)(\mi+k)}\IEEEeqnarraynumspace\label{eq:asym_cp1}
	\end{IEEEeqnarray}
	\setcounter{mycounter}{\value{equation}}
	\setcounter{equation}{\value{mycounter}+1}
	as $T\to0$, where $B(x,y)\mathdef\tfrac{\Gamma(x)\Gamma(y)}{\Gamma(x+y)}$ is the Beta function\cite{olver10}, $\kappa\mathdef\pi\lambda d^2(\md/\mi)^{2/\alpha}$ and $C_k$ is given by \eqref{eq:asym_cp2} at the top of the page.
\end{corollary}

Note that \eqref{eq:asym_cp1} is a closed-form expression, i.e., it does neither contain an improper integral nor higher-order derivatives. The integral in \eqref{eq:asym_cp2} can be solved using standard numerical software. For the special case $\md=\mi=1$ (Rayleigh fading model) and $\alpha=4$, we obtain $C_0=2 +2^{-3/2}\log(6 - 4\sqrt{2})-2^{-1/2}\log(2+\sqrt{2})\approx0.753$ and \eqref{eq:asym_cp1} then reduces to
\begin{IEEEeqnarray}{rCl}
	\cp\sim1-\kappa\,T^{1/2}\frac{\pi}{2}\left(1-\frac{3}{4}\,0.753\right)\quad\text{as }T\to0.
\end{IEEEeqnarray}

Figure~\ref{fig:asym_ps} shows the outage probability for the exact, NC, and FC model in the small outage regime for $\md=4$, $\mi=1.5$ and $\alpha=3.5$. Also shown is the asymptotic expression from \eqref{eq:asym_cp1} of Corollary~\ref{cor:asym_cp}. For reference, we also included the asymptotic outage probability expression from\cite[(5.24)]{chopra11} for $N$-antenna MRC for the isotropic interference model
\begin{IEEEeqnarray}{rCl}
   1-\cp^{\text{blind}}(N)\sim\kappa\,T^{2/\alpha}\frac{\Gamma(\mi+\tfrac{2}{\alpha})\,\Gamma(N\md-\tfrac{2}{\alpha})}{\Gamma(\mi)\,\Gamma(N\md)}.\IEEEeqnarraynumspace\label{eq:chopra}
\end{IEEEeqnarray}

We refer to \eqref{eq:chopra} as the asymptotic outage probability for {\it interference-blind} MRC, since in the isotropic interference model the MRC combining weights in\cite{chopra11} depend only on the fading gains of the desired link, cf.~\cite[Sec.~5.5.2]{chopra11}. First, it can be seen that the semi-numerical approach discussed in Section~\ref{sec:diff} accurately reflects the performance also in the low outage regime. Furthermore, the asymptotic expression in \eqref{eq:chopra} for interference-blind MRC corresponds to the outage probability for the FC model as $T\to0$. This is intuitively clear as the combining weights for interference-blind MRC do not take into account varying interference power across antennas; as a result, the combining is performed presuming identical interference power at all antennas, which corresponds to the FC model. 

We further observe that the NC model cannot capture the true diversity order as the diversity that can be harvested is significantly overestimated. A similar insight was obtained in\cite{tanbourgi13_2} for the case of Rayleigh fading links.

\begin{remark}\label{rem:mrc_gain}
	The first term in \eqref{eq:asym_cp1} corresponds to the asymptotic success probability for single-antenna receivers, which was derived in\cite{ganti11}. Hence, the second term in \eqref{eq:asym_cp1} characterizes the success probability gain due to dual-branch MRC.
\end{remark}

By Remark~\ref{rem:mrc_gain}, the outage probability for the above special case $\md=\mi=1$ and $\alpha=4$ is hence reduced by $56.2$\% when switching from single-antenna to dual-branch MRC in the asymptotic regime. We next extend this observation to the case of different $m$ and $\alpha$. Fig.~\ref{fig:rel_gain_single_antenna} shows the relative reduction in outage probability in the asymptotic regime when switching from a single-antenna system to dual-branch MRC. The relative reduction is denoted by $\Delta_{\text{MRC-SA}}$ and can be obtained by making use of Remark~\ref{rem:mrc_gain}. As expected, decreasing the per-antenna $\sinr$ variance through increasing either the path loss exponent $\alpha$ or the Nakagami parameter $m$ reduces the relative improvement of MRC. For typical path loss exponents $3<\alpha<6$, the relative improvement is $20\%<\Delta_{\text{MRC-SA}}<40\%$ for large $m$, and $40\%<\Delta_{\text{MRC-SA}}<70\%$ for small $m$ (close to Rayleigh fading).

\subsection{Comparison with other Diversity Combining Techniques}\label{sec:other_div_com}

\begin{figure*}[!t]
	\centerline{\subfloat[Success probability]{\includegraphics[width=0.474\textwidth]{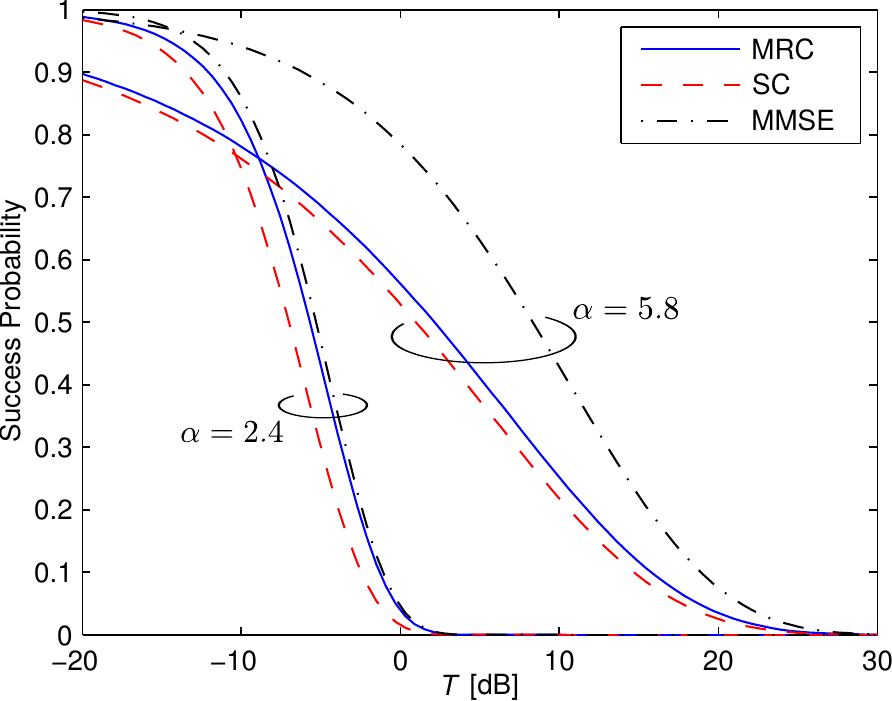}
	\label{fig:comp_pc_mrc_se_mmse}}
	\hfil
	\subfloat[Relative Diversity Gain]{\includegraphics[width=0.48\textwidth]{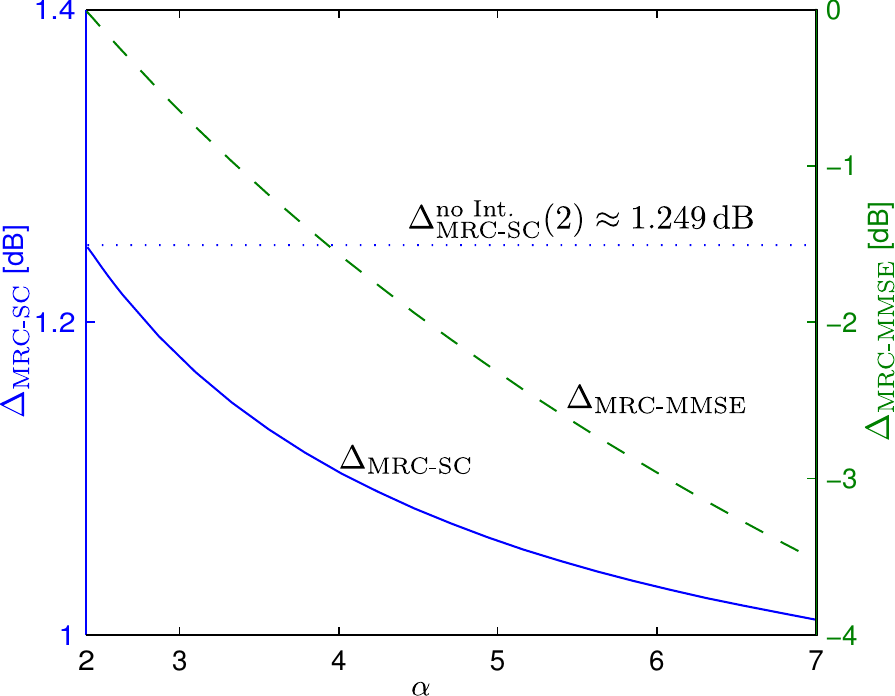}
	\label{fig:div_gain_mrc_se_mmse}}}
	\caption{(a) Success probability vs. $\sinr$-threshold $T$ for different $\alpha$. Parameters are: $\lambda=10^{-3}$, $m=1$, $d=15$, $\snr=\infty$.}
	\vspace*{4pt}
\end{figure*}

Besides MRC there also exist other diversity combining techniques, which differ in both performance and implementation complexity. The latter is generally dictated by the system design and hardware requirements, and hence does not change with the radio environment. This is, however, not true for the expected performance as different set of assumptions about the radio environment may lead to a significantly different performance prediction. In order to better understand the performance-complexity trade-offs involved in diversity combining techniques, it is therefore essential to study them under more realistic model assumptions. In the following, we will compare the expected performance of MRC with two other popular schemes, namely SC and MMSE combining, under spatially correlated interference. 

In SC, only the branch with the highest instantaneous individual $\sinr$ is selected. SC therefore has a lower complexity at the cost of a lower performance compared to MRC. In\cite{haenggi12_1}, the success probability $\cpsc$ of SC under correlated interference without noise was derived for Rayleigh fading ($m=1$) as
\begin{IEEEeqnarray}{rCl}
	\cpsc&=&\sum_{n=1}^{N}(-1)^{n+1}\binom{N}{n}\exp\left(-\Delta\,T^{2/\alpha}D_n(2/\alpha)\right),\IEEEeqnarraynumspace
\end{IEEEeqnarray}
where $\Delta\mathdef\lambda\tfrac{2\pi^2}{\alpha} d^2\csc(2\pi/\alpha)$ and $D_{n}(x)\mathdef\prod_{i=1}^{n-1}(1+x/i)$ is the so-called diversity polynomial.

In MMSE combining, the combining weights are chosen so as to maximize the post-combiner $\sinr$ under knowledge of the interference autocorrelation matrix. The success probability $\cpmmse$ for MMSE combining under Rayleigh fading ($m=1$) was derived in\cite{ali10} as
\begin{IEEEeqnarray}{rCl}
	\cpmmse&=&Q\left(N,\Delta T^{2/\alpha}+\frac{d^{\alpha}T}{\snr}\right).
\end{IEEEeqnarray}

Note that similar expressions for SC and MMSE combining for the case of Nakagami fading are currently not available in the literature. Generalizing the SC and MMSE results to Nakagami fading is beyond the scope of this contribution and is left for possible future work.

Figure~\ref{fig:comp_pc_mrc_se_mmse} compares the success probability of MRC, SC and MMSE combining for $\md=\mi=1$ (Rayleigh fading) and different $\alpha$. The performance of MRC is sandwiched by SC on the lower end and MMSE combining on the upper end as expected. Interestingly, the success probability for MRC and MMSE combining become similar as $\alpha$ decreases. This means that for small $\alpha$ almost no benefit can be harvested from estimating the interference and adapting the combining weights accordingly, compared to simply treating interference as white noise. However, such a trend is not observed for SC, where the horizontal width of the success probability gap varies no more than about 1.2 dB over a wide range of $T$ independent of $\alpha$. These observations are further elucidated in Fig.~\ref{fig:div_gain_mrc_se_mmse}, which shows the relative diversity gains $\Delta_{\text{MRC-SC}}\mathdef\mathbb{E}[\sinrmrc]/\mathbb{E}[\sinr_{\text{SC}}]$ and $\Delta_{\text{MRC-MMSE}}\mathdef\mathbb{E}[\sinrmrc]/\mathbb{E}[\sinr_{\text{MMSE}}]$ over $\alpha$ for the respective combining methods. The expectations in $\Delta_{\text{MRC-SC}}$ and $\Delta_{\text{MRC-MMSE}}$ can be obtained using the relation $\mathbb{E}[\sinr]=\int_0^{\infty}\mathbb{P}(\sinr>T)\,\mathrm dT$. 

It can be seen that $\Delta_{\text{MRC-MMSE}}$ (in dB) grows almost linearly in $\alpha$. Relative to SC, the diversity gain of MRC is roughly above 1~dB for practically relevant path loss exponents. This gain over SC, however, is always smaller than in the well-studied interference-free case. In the latter, the relative diversity gain for Rayleigh fading ($m=1$) can be written for arbitrary $N$ in terms of the harmonic series as $\Delta_{\text{MRC-SC}}^{\text{no Int.}}(N)\mathdef N\,(\sum_{n=1}^{N}1/n)^{-1}$\cite{goldsmith05}, which yields $\Delta_{\text{MRC-SC}}^{\text{no Int.}}(2)\approx1.249$ dB for the dual-branch case. The fact that $\Delta_{\text{MRC-SC}}<\Delta_{\text{MRC-SC}}^{\text{no Int.}}(N)$ for arbitrary $N$ and $\alpha>2$ can be easily verified using Jensen's inequality\cite{feller71}
\begin{IEEEeqnarray}{rCl}
	\Delta_{\text{MRC-SC}}
&=&\frac{\mathbb{E}\left[\frac{\mathsf{g}_{1}}{\mathsf{I}_{1}}+\ldots+\frac{\mathsf{g}_{N}}{\mathsf{I}_{N}}\right]}{\mathbb{E}_{\mathsf{g}_{1}\ldots\mathsf{g}_{N}}\left[\mathbb{E}_{\mathsf{I}_{1}\ldots\mathsf{I}_{N}}\left[\max\left\{\frac{\mathsf{g}_{1}}{\mathsf{I}_{1}},\ldots,\frac{\mathsf{g}_{N}}{\mathsf{I}_{N}}\right\}\right]\right]}\IEEEeqnarraynumspace\IEEEnonumber\\
&\overset{\mathrm{(a)}}{\leq}&\frac{N\mathbb{E}\left[\frac{\mathsf{g}}{\mathsf{I}}\right]}{\mathbb{E}_{\mathsf{g}_{1}\ldots\mathsf{g}_{N}}\left[\max\left\{\mathbb{E}_{\mathsf{I}_{1}}\left[\frac{\mathsf{g}_{1}}{\mathsf{I}_{1}}\right],\ldots,\mathbb{E}_{\mathsf{I}_{N}}\left[\frac{\mathsf{g}_{N}}{\mathsf{I}_{N}}\right]\right\}\right]}\IEEEeqnarraynumspace\IEEEnonumber\\
&\overset{\mathrm{(b)}}{=}&\frac{\mathbb{E}_{\mathsf{I}}\left[\mathsf{I}^{-1}\right]\,N\,\mathbb{E}[\mathsf{g}]}{\mathbb{E}_{\mathsf{I}}\left[\mathsf{I}^{-1}\right]\,\mathbb{E}_{\mathsf{g}_{1}\ldots\mathsf{g}_{N}}\left[\max\left\{\mathsf{g}_{1},\ldots,\mathsf{g}_{N}\right\}\right]}\IEEEeqnarraynumspace\IEEEnonumber\\
&=&\Delta_{\text{MRC-SC}}^{\text{no Int.}}(N),
\end{IEEEeqnarray}

where (a) follows from the fact that $\mathsf{g}_{1}/\mathsf{I}_{1},\ldots,\mathsf{g}_{N}/\mathsf{I}_{N}$, and hence the $\max$ function are convex in $\mathsf{I}_{1},\ldots,\mathsf{I}_{N}$, and by Jensen's inequality, (b) follows from the $\mathsf{I}_{n}$ being identically distributed. Note that the inequality in (a) applies not only to the Rayleigh fading case. 

Interestingly, we have $\Delta_{\text{MRC-SC}}\to\Delta_{\text{MRC-SC}}^{\text{no Int.}}(2)\approx1.249$ dB as $\alpha\to2$. This can be explained by the fact that as $\alpha\to2$, the $\mathsf{I}_{n}$ degenerate to $\mathsf{I}_{n}\equiv\infty$ almost surely\cite{HaenggiBook}. For a degenerate random variable, Jensen's inequality becomes an equality.

\subsection{Transmission Capacity for Dual-Branch MRC Receivers}\label{sec:tc}

In decentralized wireless networks such as \textit{ad hoc} networks, it is desirable to know the maximum number of local transmissions that can take place simultaneously subject to a quality of service constraint. Such a local throughput metric was introduced in\cite{weber10} under the term \textit{transmission capacity}, which is defined as
\begin{IEEEeqnarray}{c}
	c(\epsilon)\mathdef \lambda(\epsilon)\,(1-\epsilon),\quad 0\leq\epsilon\leq1,
\end{IEEEeqnarray}
where $\epsilon$ is the (target) outage probability and $\lambda(\epsilon)$ is the maximum allowable density of simultaneously active transmitters such that the success probability is at least $1-\epsilon$. We refer to\cite{weber10,HaenggiBook} for further elaborations on this metric. Since the success probability is in general monotonic in $\lambda$, $\lambda(\epsilon)$ can be obtained by (numerically) solving $\cp$ in \eqref{eq:pc_theorem} for $\lambda$, yielding the transmission capacity under dual-branch MRC.

\begin{figure}[t!]
	\centering
  \includegraphics[width=.48\textwidth]{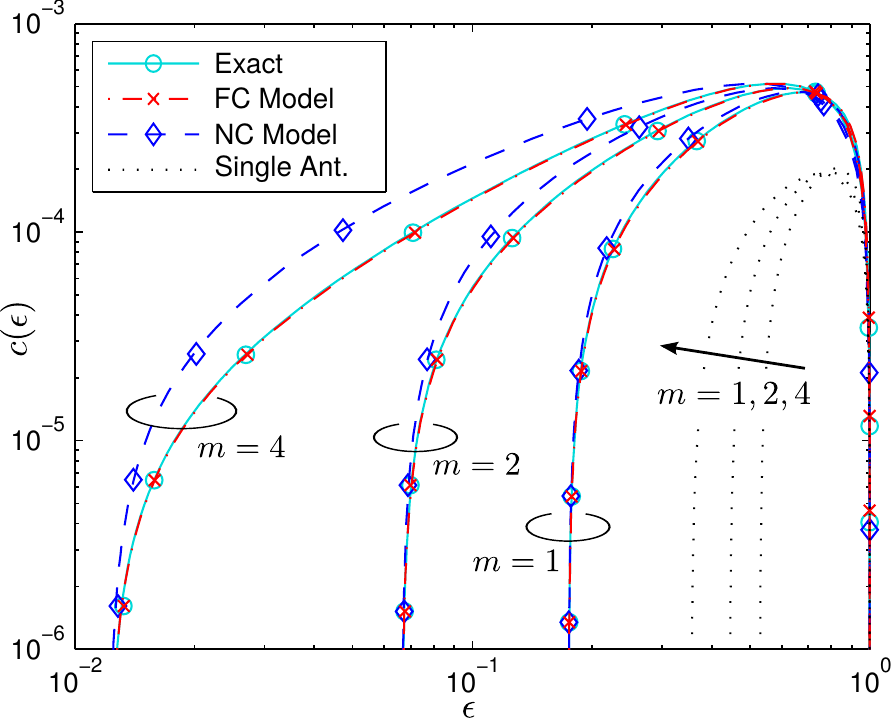}  
  \caption{Transmission capacity $c(\epsilon)$ vs. target outage probability $\epsilon$ for different $\md=\mi=m$ (identical Nakagami fading). Parameters are: $T=3$, $d=10$, $\alpha=4$, $\snr=6$~dB. Marks represent simulation results.}\label{fig:tc}
\end{figure}

Figure~\ref{fig:tc} shows the transmission capacity under dual-branch MRC for different $m$ (identical Nakagami fading). Consistent with the observations made in Section~\ref{sec:simple_models}, the FC and NC models yield a slightly pessimistic and a significantly optimistic result, respectively. Interestingly, while the accuracy loss in the NC model scales with the Nakagami fading parameter as expected, the transmission capacity gap between the FC and the exact models is fairly small even for $m=1$. The transmission capacity for the single-antenna case is also shown for reference. They were computed using \eqref{eq:prob_nc} and setting $N=1$. As can be seen, tremendous gains can be obtained when switching from single-antenna to dual-antenna MRC. These gains increase with the Nakagami fading parameter.

\section{Conclusion and Future Work}
In this paper, we developed a theoretical framework to analyze the post-combiner $\sinr$ for MRC under interference-induced correlation, independent Nakagami channel fading and receiver noise. An exact expression for the success probability was derived in semi-closed form for the dual-branch case. Our analysis concretely demonstrated that while ignoring interference correlation, thereby overestimating the true diversity, may result in significantly misleading results, assuming the same interference levels at all the antennas, thereby underestimating the true diversity, provides reasonable results when the Nakagami fading parameter of the interfering links is greater than one and/or the path loss exponent is large. In such scenarios, the frequently used full-correlation model may hence be justified. It was also shown that treating interference not as white noise through MMSE combining does not provide substantial diversity gains compared to MRC when the path loss exponent is small. Also, the gain of MRC over SC in terms of diversity gain is smaller when interference dominates noise, and this gain decays with the path loss exponent. It is important to mention that the net performance of MRC, e.g., the average rate, will also depend on the temporal correlation of the fading channel as well as of the interference. Since the locations of interferers are likely to not change significantly within consecutive transmission attempts, positive temporal interference correlation will affect the joint statistics of the $\sinr$ over time\cite{Haenggi14twc}.

This work has numerous extensions. Since our analysis was limited to the dual-branch case, a useful research direction would be to extend this framework to more than two receive antennas. The approach used in this work, namely first conditioning on the $\sinr$ in one branch and applying elementary point process theory results, does not look promising for this purpose and hence, a different approach that similarly benefits from basic stochastic geometry tools would be mandatory. Besides MRC, the performance of other combining techniques should be studied under spatially-correlated interference and fairly general fading channels, e.g., Nakagami channel fading. For instance, one concrete problem in this direction could be to study the performance of MMSE under correlation and Nakagami fading, using tools developed in~\cite{ali10} and this paper. In addition to correlation in space, the impact of temporal interference correlation\cite{Haenggi14twc} on diversity combining techniques may further be of interest to develop robust re-transmission and/or space-time coding schemes for multi-antenna systems. Another rich direction of work is to extend this framework to account for multiple antennas at the transmitter.

\appendix

\subsection{Derivation of the \texorpdfstring{$\sinr$}{SINR} for interference-aware MRC}\label{sec:mrc_model}
In this section, we rigorously derive the post-combiner $\sinr$ expression of \eqref{eq:sir_general} for an arbitrary number of antennas. For given realizations of the point process $\Phi$ and of the channel fading gains, the interference-plus-noise corrupted time-discrete signal at the $n\th$ antenna can be written as
\begin{IEEEeqnarray}{rCl}
	r_{n} = \sqrt{g_{n}}\,e^{j\theta_n}s_0+\sum\limits_{i=1}^{\infty}\sqrt{h_{n,i}}\,e^{j\phi_{n,i}}\left(\frac{d}{\|x_i\|}\right)^{\alpha/2}\hspace{-.05cm}s_{i}+w,\IEEEeqnarraynumspace
\end{IEEEeqnarray}
where $s_0$ is the desired signal, $s_{i>0}$ is the signal transmitted by the $i\th$ interferer, $\theta_n$ and $\phi_{n,i}$ describe the phase rotation on the link from desired transmitter (respectively from the $i\th$ interferer) to the $n\th$ antenna of the considered user, and $w$ is AWGN. We assume that, within the duration of one transmission slot,
\begin{enumerate}[i)]
	\item $\mathbb{E}[s_i]=0$ and $\mathbb{E}[|s_i|^2]=1$ for all $i\in\mathbb{N}_{0}$, e.g., MPSK,
	\item $\mathbb{E}[s_i s_k^{\ast}]=0$ for all $i\neq k$, $i,k\in\mathbb{N}_{0}$, 
	\item average receiver noise power is $\mathbb{E}[|w|^2]=1/\snr$.
\end{enumerate}

We further assume that the receiver can perfectly estimate the instantaneous channel $\sqrt{g_{n}}\,e^{j\theta_n}$ as well as the current interference-plus-noise signal variance (or equivalently, interference-plus-noise power) in one slot, which is given by
\begin{IEEEeqnarray}{rCl}
	&&\text{Var}_{s_1,\ldots,s_2}\left[\sum\limits_{i=1}^{\infty}\sqrt{h_{n,i}}\,e^{j\phi_{n,i}}\left(\tfrac{d}{\|x_i\|}\right)^{\alpha/2}s_{i}+w\right]\IEEEnonumber\\
	&&\quad=d^{\alpha}\sum\limits_{i=1}^{\infty}h_{n,i}\|x_i\|^{-\alpha}\,\mathbb{E}[|s_i|^2]+\mathbb{E}[|w|^2]\IEEEnonumber\\
	&&\qquad+d^{\alpha}\sum\limits_{i\neq k}\sqrt{\frac{h_{n,i}h_{n,k}}{(\|x_i\|\,\|x_k\|)^{\alpha}}}\,e^{j\phi_{n,i}-j\phi_{n,k}}\mathbb{E}[s_i s_k^{\ast}]\IEEEnonumber\\
	&&\quad=\underbrace{d^{\alpha}\sum\limits_{i=1}^{\infty}h_{n,i}\|x_i\|^{-\alpha}}_{I_n}+\frac{1}{\snr},
\end{IEEEeqnarray}
where the last step follows from assumptions i) -- iii). Estimation of the interference-plus-noise signal variance can be performed using techniques such as those proposed in\cite{benedict67,pauluzzi00}, for instance during the channel training period after having determined $\sqrt{g_{n}}\,e^{j\theta_n}$. Note that one can assume $I_n$ to be finite since $\mathsf{I}_{n}<\infty$ almost surely when $\alpha>2$\cite{HaenggiBook}.

Under the hypothesis that interference is treated as white noise, the $\sinr$ is maximized by MRC. According to\cite{brennan03}, the MRC weight $a_n$ at the $n\th$ antenna is chosen as $a_n=s_0^{\ast}\sqrt{g_n}e^{-j\theta_n}/(I_n+\snr^{-1})$. The $\sinr$ then takes the form
\begin{IEEEeqnarray}{rCl}
	\sinr&=& \frac{\left(\sum_{n=1}^{N}a_n\,\sqrt{g_{n}}\,e^{j\theta_n}s_0\right)^2}{\text{Var}\hspace{-.05cm}\left[\sum_{n=1}^{N}a_n\left(\sum\limits_{i=1}^{\infty}\sqrt{h_{n,i}}\,e^{j\phi_{n,i}}\hspace{-.05cm}\left(\frac{d}{\|x_i\|}\right)^{\alpha/2}\hspace{-.2cm}s_{i}+w\hspace{-.05cm}\right)\right]}\IEEEnonumber\\
	&=&\frac{\left(\sum_{n=1}^{N}\frac{g_{n}}{I_n+\snr^{-1}}\right)^2}{\sum_{n=1}^{N}\frac{g_{n}(I_n+\snr^{-1})}{(I_n+\snr^{-1})^2}}=\sum\limits_{n=1}^{N}\frac{g_{n}}{I_n+\snr^{-1}}.\label{eq:sinr_realization}\IEEEeqnarraynumspace
\end{IEEEeqnarray}
De-conditioning \eqref{eq:sinr_realization} upon $\Phi$ and the channel fading gains, we finally obtain the (random) post-combiner $\sinr$ of \eqref{eq:sir_general}.  
\subsection{Proof of Theorem~\ref{thm:cov_prob}}\label{proof:cov_prob}
Define the auxiliary random variable
\begin{IEEEeqnarray}{rCl}
	\mathsf{Z}\mathdef\frac{\mathsf{g}_2}{\mathsf{I}_{2}+\snr^{-1}}
\end{IEEEeqnarray}
and condition $\cp$ on the point process $\Phi$ and $\mathsf{Z}$, yielding
\begin{IEEEeqnarray}{rCl}
	\cp&=&\mathbb{E}_{\Phi,\mathsf{Z}}\Big[\mathbb{P}\left(\mathsf{g}_1\geq(T-\mathsf{Z})(\mathsf{I}_{1}+\snr^{-1})\,\big\lvert\,\Phi,\mathsf{Z}\right)\Big].\IEEEeqnarraynumspace\label{eq:proof_thm1_step1}
\end{IEEEeqnarray}
The conditional probability in \eqref{eq:proof_thm1_step1} can be written as
\begin{IEEEeqnarray}{rCl}
&&\mathbb{P}\left(\mathsf{g}_1\geq(T-\mathsf{Z})(\mathsf{I}_{1}+\snr^{-1})\,\big\lvert\,\Phi,\mathsf{Z}\right)\IEEEnonumber\IEEEeqnarraynumspace\\
&&\quad=\mathbb{E}_{\mathbf{h}_1}\Big[\mathbb{P}\left(\mathsf{g}_1\geq(T-\mathsf{Z})(\mathsf{I}_{1}+\snr^{-1})\,\big\lvert\,\Phi,\mathsf{Z},\mathbf{h}_{1}\right)\Big]\IEEEnonumber\IEEEeqnarraynumspace\\
&&\quad=\mathbb{E}_{\mathbf{h}_1}\left[\frac{1}{\Gamma(\md)}\,\Gamma\big(\md,\psi_{1}\,(\mathsf{I}_{1}+\snr^{-1})\big)\right],\IEEEeqnarraynumspace\label{eq:proof_thm1_step2}
\end{IEEEeqnarray}
where we have performed the substitution $\psi_{1}\mathdef(T-\mathsf{Z})^{+}\md$. Using the fact that $\Gamma(a,x)/\Gamma(a)=\sum_{k=0}^{a-1}x^ke^{-x}/k!$ for integer $a$ \cite{olver10}, we rewrite \eqref{eq:proof_thm1_step2} for integer $\md$ as
\begin{IEEEeqnarray}{rCl}
	&&\mathbb{E}_{\mathbf{h}_1}\left[\frac{1}{\Gamma(\md)}\,\Gamma\big(\md,\psi_{1}\,(\mathsf{I}_{1}+\snr^{-1})\big)\right]\IEEEnonumber\IEEEeqnarraynumspace\\
	&&\quad=\mathbb{E}_{\mathbf{h}_1}\bigg[\sum\limits_{k=0}^{\md-1}\frac{1}{k!}\,\big(\underbrace{\psi_{1}\,(\mathsf{I}_{1}+\snr^{-1})}_{\mathdef\mathsf{Y}}\big)^{k}e^{-\psi_{1}\,(\mathsf{I}_{1}+\snr^{-1})}\bigg]\IEEEeqnarraynumspace\IEEEnonumber\\
	&&\quad=\sum\limits_{k=0}^{\md-1}\frac{(-1)^{k}}{k!}\,\mathbb{E}_{\mathsf{Y}}\left[(-1)^{k}\mathsf{Y}^{k}e^{-\mathsf{Y}}\right]\IEEEnonumber\IEEEeqnarraynumspace\\
	&&\quad=\sum\limits_{k=0}^{\md-1}\frac{(-1)^{k}}{k!}\frac{\partial^{k}}{\partial s^{k}}\Big[\mathcal{L}_{\mathsf{Y}}(s)\Big]_{s=1}.\IEEEeqnarraynumspace\label{eq:proof_thm1_step3}
\end{IEEEeqnarray}
The Laplace transform $\mathcal{L}_{\mathsf{Y}}(s)$ can be obtain as
\begin{IEEEeqnarray}{rCl}
	\mathcal{L}_{\mathsf{Y}}(s) &=& \mathbb{E}_{\mathrm h_1}\left[\exp\left(-s\psi_{1}\left(\snr^{-1}+\sum\limits_{\mathsf{x}_{i}\in\Phi}\mathsf{h}_{1,i}\|\mathsf{x}_{i}\|^{-\alpha}\right)\right)\right]\IEEEnonumber\\
		&\overset{\text{(a)}}{=}&e^{-\frac{s\psi_{1}}{\snr}} \prod\limits_{\mathsf{x}_{i}\in\Phi}\mathbb{E}_{\mathsf{h}_{1}}\big[e^{-s\psi_{1}d^{\alpha}\mathsf{h}_{1}\|\mathsf{x}_{i}\|^{-\alpha}}\big]
,\IEEEeqnarraynumspace\label{eq:proof_thm1_step4}
\end{IEEEeqnarray}
where (a) follows from the i.i.d. fading property. 
Hence, combining \eqref{eq:proof_thm1_step3} and \eqref{eq:proof_thm1_step4}, we can rewrite \eqref{eq:proof_thm1_step2} as
\begin{IEEEeqnarray}{c}
	\sum\limits_{k=0}^{\md-1}\hspace{-.12cm}\frac{(-1)^{k}}{k!}\frac{\partial^{k}}{\partial s^{k}}\hspace{-.08cm}\left[e^{-\frac{s\psi_{1}}{\snr}}\hspace{-.1cm}\prod\limits_{\mathsf{x}_{i}\in\Phi}\hspace{-.05cm}\mathbb{E}_{\mathsf{h}_{1}}\big[e^{-s\psi_{1}d^{\alpha}\mathsf{h}_{1}\|\mathsf{x}_{i}\|^{-\alpha}}\big]\right]_{s=1}\hspace{-.4cm}.\label{eq:proof_thm1_step5}\IEEEeqnarraynumspace
\end{IEEEeqnarray}

For averaging over $\mathsf{Z}$ conditional on $\Phi$, we first calculate the conditional PDF
\begin{IEEEeqnarray}{rCl}
	&&\frac{\partial}{\partial z}\,\mathbb{P}\left(\mathsf{Z}\leq z\,\lvert\,\Phi\right)\IEEEnonumber\\
	&&=\frac{\partial}{\partial z}\mathbb{E}_{\mathrm h_2}\big[\mathbb{P}\left(\mathsf{g}_2\leq z\,\left(\mathsf{I}_{2}+\snr^{-1}\right)\left\lvert\right.\Phi\right)\big]\IEEEnonumber\\
	&&\quad=\frac{\partial}{\partial z}\mathbb{E}_{\mathrm h_2}\left[\frac{1}{\Gamma(\md)}\,\gamma\left(\md,z\md\,\left(\mathsf{I}_{2}+\snr^{-1}\right)\right)\right]\IEEEnonumber\\
	&&\quad\overset{\text{(a)}}{=} \mathbb{E}_{\mathrm h_2}\left[\frac{1}{\Gamma(\md)}\,\frac{\partial}{\partial z}\gamma\left(\md,\psi_{2}\left(\mathsf{I}_{2}+\snr^{-1}\right)\right)\right]\IEEEnonumber\\
	&&\quad\overset{\text{(b)}}{=}\frac{z^{-1}}{\Gamma(\md)}\,\mathbb{E}_{\mathrm h_2}\left[\big(\psi_{2}\left(\mathsf{I}_{2}+\snr^{-1}\right)\big)^{\md} e^{-\psi_{2}\,(\mathsf{I}_{2}+\snr^{-1})}\right]\IEEEnonumber\\
	&&\quad\overset{\text{(c)}}{=}\frac{(-1)^{\md}}{z\,\Gamma(\md)}\frac{\partial^{\md}}{\partial t^{\md}}\hspace{-.1cm}\left[e^{-\frac{t\psi_{2}}{\snr}}\hspace{-.05cm}\prod\limits_{\mathsf{x}_{i}\in\Phi}\hspace{-.02cm}\mathbb{E}_{\mathsf{h}_{2}}\hspace{-.03cm}\big[e^{-t\psi_2d^{\alpha}\mathsf{h}_{2}\|\mathsf{x}_{i}\|^{-\alpha}}\big]\right]_{t=1}\hspace{-.35cm},\IEEEeqnarraynumspace\label{eq:proof_thm1_step6}
\end{IEEEeqnarray}
where (a) follows from the dominated convergence theorem \cite{olver10} and from substituting $\psi_{2}\mathdef z\md$, (b) is obtained using the relation $\partial/\partial x\,\gamma(a,x)=x^{a-1}e^{-x}$ \cite{olver10} where $\gamma(a,x)\mathdef\int_{0}^{x}t^{a-1}e^{-t}\mathrm dt$ is the lower incomplete Gamma function, and (c) follows from applying the same technique for obtaining \eqref{eq:proof_thm1_step3} and \eqref{eq:proof_thm1_step4}. Substituting \eqref{eq:proof_thm1_step5} and \eqref{eq:proof_thm1_step6} into \eqref{eq:proof_thm1_step1}, $\cp$ can be written as

\abovedisplayskip=5pt
\belowdisplayskip=5pt

\begin{IEEEeqnarray}{Rl}
&\mathbb{E}_{\Phi,\mathsf{Z}}\hspace{-.02cm}\Bigg[\hspace{-.02cm}\sum\limits_{k=0}^{\md-1}\hspace{-.05cm}\frac{(-1)^{k}}{k!}\frac{\partial^{k}}{\partial s^{k}}\hspace{-.02cm}\bigg[\hspace{-.02cm}e^{-\frac{s\psi_{1}}{\snr}}\hspace{-.12cm}\prod\limits_{\mathsf{x}_{i}\in\Phi}\hspace{-.1cm}\mathbb{E}_{\mathsf{h}_{1}}\hspace{-.07cm}\Big[e^{-s\psi_{1}\frac{d^{\alpha}}{\|\mathsf{x}_i\|^{\alpha}}\mathsf{h}_{1}}\Big]\bigg]_{s=1}\hspace{-.04cm}\Bigg]\IEEEnonumber\\
	&=\sum\limits_{k=0}^{\md-1}\hspace{-.1cm}\frac{(-1)^{k}}{k!}\mathbb{E}_{\Phi,\mathsf{Z}}\Bigg[\frac{\partial^{k}}{\partial s^{k}}\hspace{-.05cm}\bigg[e^{-\frac{s\psi_{1}}{\snr}} \hspace{-.07cm}\prod\limits_{\mathsf{x}_{i}\in\Phi}\hspace{-.1cm}\mathbb{E}_{\mathsf{h}_{1}} \hspace{-.05cm}\Big[e^{-s\psi_{1}\frac{d^{\alpha}}{\|\mathsf{x}_i\|^{\alpha}}\mathsf{h}_{1}}\Big]\bigg]_{s=1}\Bigg]\IEEEnonumber\\
	&=\sum\limits_{k=0}^{\md-1}\frac{(-1)^{k}}{k!}\int_{0}^{\infty}\frac{(-1)^{\md}}{z\,\Gamma(\md)}\IEEEnonumber\\
&\qquad\times\mathbb{E}_{\Phi}\Bigg[\frac{\partial^{k}}{\partial s^{k}}\bigg[e^{-\frac{s\psi_{1}}{\snr}}\prod\limits_{\mathsf{x}_{i}\in\Phi}\hspace{-.1cm}\mathbb{E}_{\mathsf{h}_{1}}\Big[e^{-s\psi_{1}\frac{d^{\alpha}}{\|\mathsf{x}_i\|^{\alpha}}\mathsf{h}_{1}}\Big]\bigg]_{s=1}\IEEEnonumber\\
	&\qquad\qquad\times\frac{\partial^{\md}}{\partial t^{\md}}\bigg[e^{-\frac{t\psi_{2}}{\snr}} \hspace{-.1cm}\prod\limits_{\mathsf{x}_{i}\in\Phi}\hspace{-.1cm}\mathbb{E}_{\mathsf{h}_{2}}\Big[e^{-t\psi_{2}\frac{d^{\alpha}}{\|\mathsf{x}_i\|^{\alpha}}\mathsf{h}_{2}}\Big]\bigg]_{t=1} \Bigg]\,\mathrm dz\IEEEnonumber\\
	&\overset{\text{(a)}}{=}\sum\limits_{k=0}^{\md-1}\frac{(-1)^{k+\md}}{k!\,\Gamma(\md)}\int_{0}^{\infty}\hspace{-.18cm}\frac{\partial^{k}\partial^{\md}}{z\,\partial s^{k}\partial t^{\md}}\Bigg[e^{-\frac{s\psi_{1}}{\snr}-\frac{t\psi_{2}}{\snr}}\IEEEnonumber\\ 	&\qquad\times\mathbb{E}_{\Phi}\hspace{-.02cm}\bigg[\hspace{-.05cm}\prod\limits_{\mathsf{x}_{i}\in\Phi}\hspace{-.1cm}\mathbb{E}_{\mathsf{h}_{1}}\hspace{-.07cm}\left[e^{-s\psi_{1}\frac{d^{\alpha}}{\|\mathsf{x}_i\|^{\alpha}}\mathsf{h}_{1}}\right]\,\mathbb{E}_{\mathsf{h}_{2}}\hspace{-.07cm}\left[e^{-t\psi_{2}\frac{d^{\alpha}}{\|\mathsf{x}_i\|^{\alpha}}\mathsf{h}_{2}}\right]\hspace{-.03cm}\bigg]\hspace{-.03cm}\Bigg]_{\substack{s=1\\ t=1}}\hspace{-.25cm}\mathrm dz\IEEEnonumber\\
	&\overset{\text{(b)}}{=}\sum\limits_{k=0}^{\md-1}\frac{(-1)^{k+\md}}{k!\,\Gamma(\md)}\IEEEnonumber\\
	&\qquad\times\int_{0}^{\infty}z^{-1}\frac{\partial^{k}\partial^{\md}}{\partial s^{k}\partial t^{\md}}\bigg[e^{-\frac{s\psi_{1}}{\snr}-\frac{t\psi_{2}}{\snr}-\pi\lambda\mathcal{A}(z,s,t)}\bigg]_{\substack{s=1\\ t=1}}\hspace{-.15cm}\mathrm dz,\label{eq:proof_thm1_step7}
\end{IEEEeqnarray}
where (a) follows from the dominated convergence theorem \cite{olver10} and (b) follows from the PGFL for PPPs \cite{stoyan95,HaenggiBook}, where
\begin{IEEEeqnarray}{rCl}
	\mathcal{A}(z,s,t)=\hspace{-.05cm}\int_{0}^{\infty}\hspace{-.27cm}2r\left(\hspace{-.05cm}1-\mathbb{E}_{\mathsf{h}_{1}\hspace{-.03cm},\mathsf{h}_{2}}\hspace{-.13cm}\left[e^{-r^{-\alpha}d^{\alpha}(s\psi_{1}\mathsf{h}_{1}+t\psi_{2}\mathsf{h}_{2})}\hspace{-.05cm}\right]\right)\mathrm dr.\label{eq:poisson_mean1}\IEEEeqnarraynumspace
\end{IEEEeqnarray}
Using the same approach as in \cite[Chap.~3.2]{ganti09}, \eqref{eq:poisson_mean1} yields
\begin{IEEEeqnarray}{rCl}
	\mathcal{A}(z,s,t)=d^{2}\,\Gamma(1-2/\alpha)\,\mathbb{E}_{\mathsf{h}_{1},\mathsf{h}_{2}}\hspace{-.10cm}\left[(s\psi_{1}\mathsf{h}_{1}+t\psi_{2}\mathsf{h}_{2})^{2/\alpha}\right].\label{eq:poisson_mean2}\IEEEeqnarraynumspace
\end{IEEEeqnarray}
For $z\geq T$, $\psi_{1}=0$ and \eqref{eq:poisson_mean2} reduces to
\begin{IEEEeqnarray}{rCl}
	\mathcal{A}(z,s,t)= (\md zt)^{2/\alpha}\,d^{2}\,\Gamma(1-2/\alpha)\,\frac{\Gamma(2/\alpha+\mi)}{\mi^{2/\alpha}\Gamma(\mi)},\IEEEeqnarraynumspace
\end{IEEEeqnarray}
since $\mathbb{E}[\mathsf{h}_{2}^{2/\alpha}]=\mi^{-2/\alpha}\Gamma(2/\alpha+\mi)/\Gamma(\mi)$. When $0\leq z\leq T$, we invoke a fractional moment result from \cite{wolfe75} to obtain
\begin{IEEEeqnarray}{rCl}
	&&\mathbb{E}_{\mathsf{h}_{1},\mathsf{h}_{2}}\big[(\underbrace{s\psi_{1}\mathsf{h}_{1}+t\psi_{2}\mathsf{h}_{2}}_{\mathsf{H}})^{2/\alpha}\big]=\frac{2/\alpha}{\Gamma(1-\tfrac{2}{\alpha})}\int_{0}^{\infty}\frac{1-\mathcal{L}_{\mathsf{H}}(u)}{u^{1+2/\alpha}}\,\mathrm du\IEEEnonumber\\
	&&\qquad\overset{\text{(a)}}{=}s^{2/\alpha}(T-z)^{2/\alpha}\left(\tfrac{\md}{\mi}\right)^{2/\alpha}\Gamma(2/\alpha+2\mi)\IEEEnonumber\\
	&&\qquad\qquad\times{}_{2}\mathbf{F}_{1}\left(-2/\alpha,\mi,2\mi,1-\tfrac{zt}{(T-z)s}\right),\IEEEeqnarraynumspace\label{eq:frac_mom_H}
\end{IEEEeqnarray}
where ${}_{2}\mathbf{F}_{1}(a,b,c,z)\mathdef{}_{2}{F}_{1}(a,b,c,z)/\Gamma(c)$ is the \textit{regularized} Gaussian hypergeometric function \cite{olver10}. (a) follows from the fact that $\mathcal{L}_{\mathsf{H}}(u)=\mathcal{L}_{s\psi_{1}\mathsf{h}_{1}}(u)\,\mathcal{L}_{t\psi_{2}\mathsf{h}_{2}}(u)=((1+us\psi_{1}/\mi)(1+ut\psi_{2}/\mi))^{-\mi}$. Hence, for $0\leq z\leq T$,
\begin{IEEEeqnarray}{rCl}
	\mathcal{A}(z,s,t) &=& s^{2/\alpha}(T-z)^{2/\alpha}\,d^{2}\,\Gamma(1-2/\alpha)\,\left(\tfrac{\md}{\mi}\right)^{2/\alpha}\IEEEnonumber\\
&&\hspace{-1.2cm}\times\Gamma(2/\alpha+2\mi)\,{}_{2}\mathbf{F}_{1}\left(-2/\alpha,\mi,2\mi,1-\tfrac{zt}{(T-z)s}\right).\IEEEeqnarraynumspace
\end{IEEEeqnarray}

\bibliographystyle{IEEEtran}
\end{document}